%% file: main.tex
\documentclass[nonacm,sigplan]{acmart}
\pdfoutput=1
\usepackage[utf8]{inputenc} 
\usepackage[T1]{fontenc}    
\usepackage{booktabs}       
\usepackage{amsfonts}       
\usepackage{nicefrac}       
\usepackage{microtype}      
\hypersetup{
    colorlinks,
    linkcolor={red!50!black},
    citecolor={blue!50!black},
    urlcolor={blue!80!black}
}

\usepackage{microtype}
\usepackage{graphicx}
\usepackage{subfig}
\usepackage{booktabs} 
\usepackage{subfig}
\usepackage{epsfig}
\usepackage{soul}
\usepackage{enumitem}
\usepackage{multirow}
\usepackage{formatting/macros}
\usepackage{balance}
\usepackage{amsmath} 
\usepackage{nccmath}
\usepackage{wrapfig}
\usepackage{algorithm}
\usepackage{algorithmic}
\usepackage{color,soul}
\usepackage{balance}
\usepackage{cancel}

\begin{document}

\author{Yassaman Ebrahimzadeh Maboud}
 \affiliation{%
    \institution{The University of British Columbia}
    \country{Vancouver, Canada}
   }
 \email{yassaman@ece.ubc.ca}

 \author{Muhammad Adnan}
 \affiliation{%
    \institution{The University of British Columbia}
    \country{Vancouver, Canada}
   }
 \email{adnan@ece.ubc.ca}

 \author{Divya Mahajan}
 \affiliation{%
   \institution{Georgia Institute of Technology}
    \country{Atlanta, USA}
   }
 \email{divya.mahajan@gatech.edu}

 \author{Prashant J. Nair}
 \affiliation{%
   \institution{The University of British Columbia}
    \country{Vancouver, Canada}
   }
 \email{prashantnair@ece.ubc.ca}

\renewcommand{\shortauthors}{Maboud et al.}

\title{Efficient Training of Recommender Models through Popularity-Based Skipping of Stale Embeddings}

\input{body/abstract}

\maketitle 
\pagestyle{plain} 

\input{body/introduction}
\input{body/background}
\input{body/slipstream}

\input{body/evaluation}

\input{body/related}

\bibliographystyle{plain}
\bibliography{ref}

\end{document}

%% file: body/abstract.tex
\begin{abstract}
Training recommendation models pose significant challenges regarding resource utilization and performance. Prior research has proposed an approach that categorizes embeddings into popular and non-popular classes to reduce the training time for recommendation models. We observe that, even among the popular embeddings, certain embeddings undergo rapid training and exhibit minimal subsequent variation, resulting in saturation. Consequently, updates to these embeddings lack any contribution to model quality.

This paper presents Slipstream, a software framework that identifies stale embeddings on the fly and skips their updates to enhance performance. This capability enables Slipstream to achieve substantial speedup, optimize CPU-GPU bandwidth usage, and eliminate unnecessary memory access. SlipStream showcases training time reductions of 2$\times$, 2.4$\times$, 1.2$\times$, and 1.175$\times$ across real-world datasets and configurations, compared to Baseline XDL, Intel-optimized DRLM, FAE, and Hotline, respectively.
 
\end{abstract}

%% file: body/introduction.tex
\section{Introduction}
Recommendation systems are pivotal in various industrial sectors, including Online Advertisement and E-commerce~\citep{netflixreco, amazonreco, facebook:ml, acun2020understanding}. These systems typically have a computationally intensive neural network and memory-intensive embedding tables. In large-scale deep-learning-based models, accessing embedding tables often incurs significant performance overheads and constrains training throughput. Prior research has explored techniques such as leveraging embedding popularity, compression, and sparsity to enhance training efficiency~further \citep{fae, recshard, compressreco, compressreco2, sparsemat}. In contrast, our study focuses on optimizing training performance by dynamically harnessing the invariability within large sets of embedding values. This paper seeks to exploit the semantic significance of popularity within recommendation models. 

\begin{figure}[t]
	\centering
	\includegraphics[width=0.8\columnwidth]{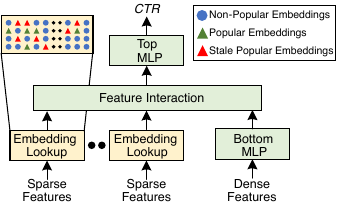}
	\caption{The Deep Learning Recommendation Model (DLRM) consists of compute-intensive Multi-Layer Perceptrons (MLPs) and memory-intensive embedding lookup operations. Due to the large embedding tables and skewed accesses, numerous embedding entries are rapidly trained and remain stagnant throughout the training process.}
	\label{fig:rec_model}
    \vspace{-0.3in}
\end{figure}

Commercial recommendation models, such as Deep Learning Recommendation Models (DLRM), utilize embedding tables to store numerical features associated with items and users~\cite {dlrm,baidu}. As depicted in Figure~\ref{fig:rec_model}, these embedding tables facilitate processing sparse features, while a bottom Multi-Layer Perceptron (MLP) handles dense features. The operations of embedding lookup and the bottom MLP are intricately linked to the feature interaction layer, which, in turn, connects to a top MLP to generate the Click Through Rate (CTR). This architecture enables efficient representation and processing of diverse feature types, facilitating accurate recommendations. Notably, embedding tables typically reach sizes of hundreds of gigabytes~\cite{baidu}, making DLRM training primarily a memory-bound operation.

In real-world scenarios, access to embeddings typically follows a highly skewed pattern. This results in a few frequently accessed embedding entries, referred to as `hot embeddings,' being accessed significantly more frequently than others. Consequently, a significant fraction of inputs are directed towards training these few embeddings, influencing training time. However, this skewed access pattern also leads to considerable input variability in these hot embeddings, enabling rapid training convergence. Once an embedding entry is fully trained, subsequent input accesses result in minimal value updates. Continuing to train these embeddings leads to unnecessary data movements and computations. 

Our paper uses these insights and proposes a software framework called Slipstream. Slipstream optimizes training efficiency by selectively updating embedding values based on data awareness. This approach enables Slipstream to reduce unnecessary computations and data movements by eliminating redundant training updates. Slipstream comprises three runtime components, instantiated one after another, to enhance the training performance and accuracy.

\begin{enumerate}
[leftmargin=0cm,itemindent=.5cm,labelwidth=\itemindent,labelsep=0cm,align=left, listparindent=0.3cm, noitemsep]
\item \textbf{Snapshot Block}: Slipstream employs a hot embedding filtering mechanism called the 'Snapshot' Block. This mechanism is designed to identify frequently accessed or `hot' embeddings that get trained rapidly. The Snapshot Block periodically captures snapshots of the `hot' embeddings during runtime. These snapshots are used to track and record the rapid training dynamics of embeddings effectively. 

Remarkably, despite the total size of embeddings in real-world datasets reaching several hundred gigabytes, the size of their hot embeddings remains relatively small, typically only a few hundred megabytes. Consequently, multiple snapshots of hot embeddings can be stored within GPU devices without imposing significant memory capacity overheads.

\item \textbf{Sampling Block}: Despite maintaining multiple temporal snapshots of 'hot' embeddings, discerning which are fully trained poses a significant challenge. An embedding entry is typically deemed fully trained when its values stabilize within a certain relative threshold. Thus, one can discard such stable embedding entries. However, iterating over these embedding table snapshots with various threshold values can incur substantial execution overheads, as observed in our experiments, with overheads increasing by 2$\times$ to 500$\times$.

To address this challenge, Slipstream implements input sampling techniques to expedite the determination of an optimal threshold. Slipstream accurately estimates the inputs associated with low-variation embeddings by iteratively sampling a subset of embeddings and adjusting thresholds. This approach significantly reduces the overheads associated with threshold determination by nearly 1000$\times$.

\item \textbf{Input Classifier Block}: After identifying stable embeddings, Slipstream optimizes training efficiency by skipping inputs associated with these embeddings for the remaining training process. This reduces communication overhead and eliminates computational overhead for these inputs.

Slipstream achieves this optimization by incorporating an input classifier block. This block selectively filters inputs accessing high-variation embeddings and trains using these inputs. Additionally, Slipstream uses a feature normalization process to attain (and significantly exceed) baseline accuracy. Thus, Slipstream enables a dynamic performance-optimized training process by leveraging the inherent popularity-related semantics within recommender models.

\end{enumerate}

\noindent Broadly, this paper has three key contributions.
\begin{itemize}[leftmargin=0cm,itemindent=.5cm,labelwidth=\itemindent,labelsep=0cm,align=left, listparindent=0.3cm, noitemsep]
\item \textbf{Contribution 1}: Our paper emphasizes the significance of leveraging semantics to identify fully trained embeddings accurately. Slipstream optimizes the training process by effectively exploiting the inherent semantic properties.
\item \textbf{Contribution 2}: This work further employs minor normalization to the baseline model to further improve accuracy. 
\item \textbf{Contribution 3}: It introduces a novel approach, Slipstream, to efficiently identify and discard fully-trained embeddings at runtime on real systems, thereby reducing computational overhead and enhancing training efficiency. 
\end{itemize}

We demonstrate that Slipstream maintains training accuracy while significantly reducing training time. Slipstream, on average, reduces the training time by 2$\times$, 2.4$\times$, 20$\%$, and 18$\%$ compared to commercial XDL~\citep{xdl}, Intel-Optimized DLRM~\citep{dlrm-intel}, FAE~\citep{fae}, and Hotline~\citep{hotline} baselines, respectively.

%% file: body/background.tex
\section{Background and Challenges}

\label{sec:background}
\subsection{A Overview of Recommendation Systems}
Deep learning-based recommendation models (DLRM) use separate learning paths for continuous and categorical features~\citep{dlrm}. Continuous features like user activity timestamps and age groups are encapsulated within dense inputs and processed by multi-layer perceptrons (MLPs) to generate condensed continuous vectors. On the other hand, categorical features, like user locations and web links, which cannot be naturally expressed in continuous form, are represented as sparse inputs. These inputs are managed by embedding tables, where each row corresponds to a potential item, and the number of columns is determined as a hyperparameter.

User-item interactions in DLRM-based systems are often sparse, as users typically interact with only a few items~\citep{fae}. To handle these sparse input vectors, an embedding lookup operation accesses arbitrary rows of an embedding table, producing a single vector per table. These sparse features then interact with dense features from the bottom MLP layer through concatenation or dot product operations. The resulting feature interactions are passed through an additional top MLP layer to generate the final ranking or Click-Through Rate (CTR) prediction. 

\subsection{Training Setup}
\label{subsec:training_setup}

Training large recommender models often involves two distributed modes: the hybrid mode and the data-parallel mode. In the hybrid mode, embeddings are stored and gathered on the CPU, while neural networks are executed on GPUs in a data-parallel fashion. This setup accommodates large embedding tables in CPU memories, effectively managing increased model sizes without necessarily scaling up the number of GPUs. However, despite its advantages, the hybrid mode often faces limitations in training throughput due to significant data transfers and reliance on the CPU's lower-bandwidth main memory. In this paper, we delve into the intricacies of the hybrid training mode. However, our insights can also be easily applied to the data-parallel GPU-only mode.

\subsection{Motivation: Data-Aware Embedding Updates}
This paper's goal is to facilitate embedding-variation-aware training at runtime. This goal stems from three key observations: the large costs associated with embedding operations, the dynamic nature of embedding access frequency, and the update patterns within embeddings over time. 

\subsubsection{Breakdown of Training Time}

\begin{figure}[t]
	\centering
	\includegraphics[width=1\columnwidth]{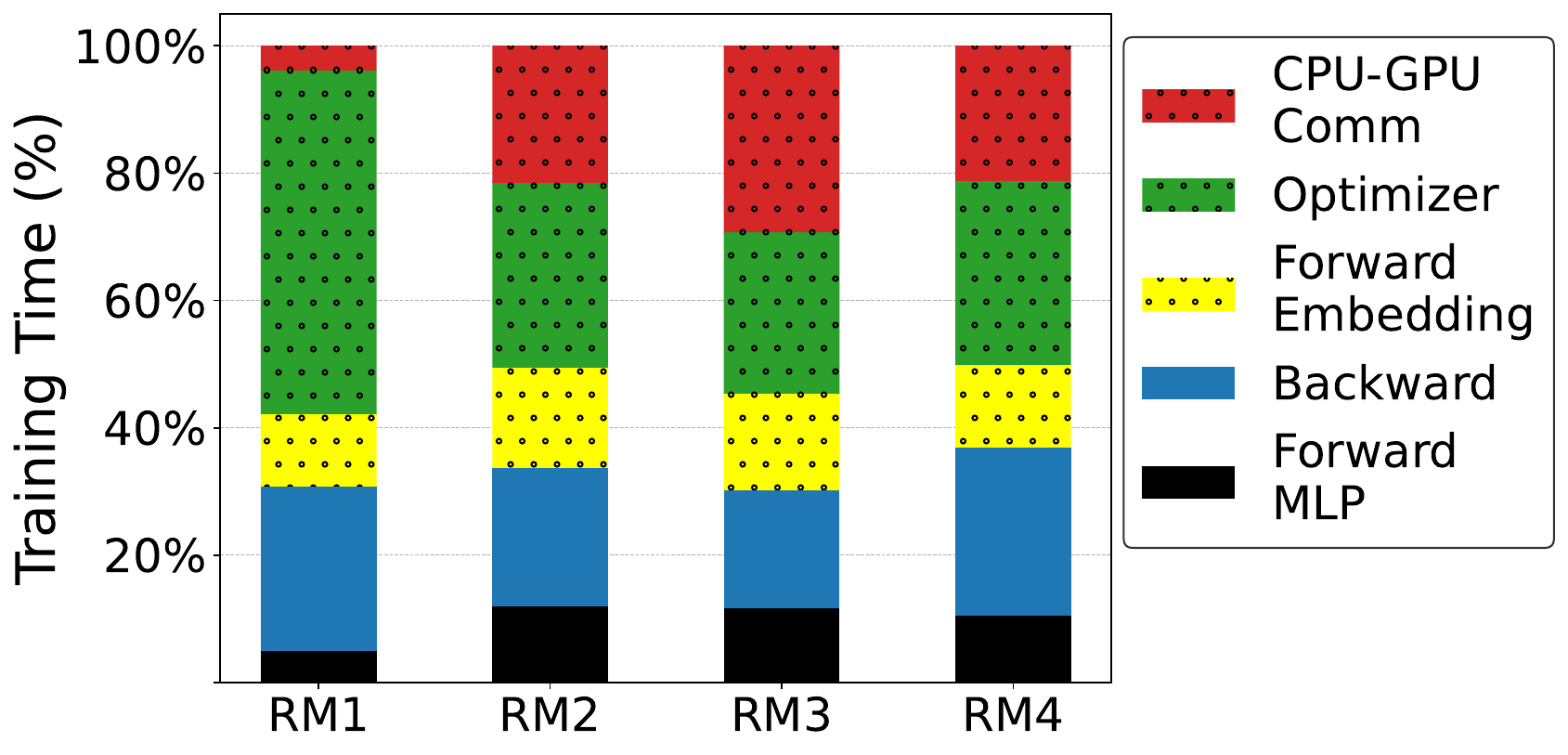}
	\caption{The breakdown of the training time for an Intel-optimized DLRM with 4-GPU in a hybrid CPU-GPU training setup. We observe that a significant fraction of the time is spent on forward embedding pass, embedding updates in the optimizer, and communication.}
	\label{fig:motivation}
\end{figure}

Figure~\ref{fig:motivation} depicts the breakdown of training time across four real-world models and datasets. Notably, it showcases that embedding operations, including embedding lookup during the forward pass, updating embeddings in the optimizer, and CPU-GPU communication, can collectively consume up to 75\% of the training time, especially in large datasets like Criteo Terabyte. This underscores the need for optimization techniques which efficiently manage embedding operations to reduce training time and significantly enhance overall model performance.

\begin{figure}[h!]

  \centering

	\begin{minipage}[t]{0.275\linewidth}
	   \centering
	   \includegraphics[width=\textwidth]{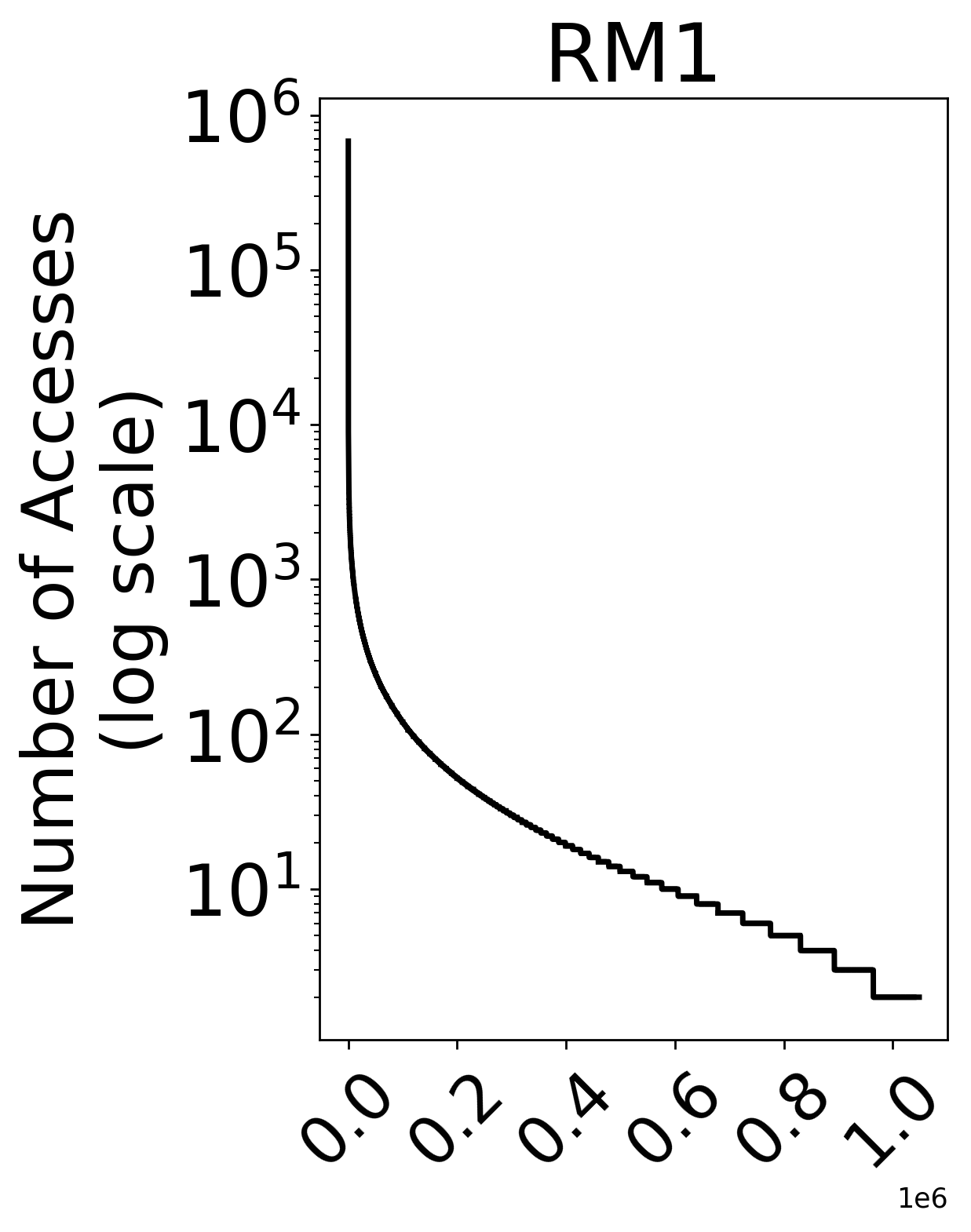}
	\end{minipage}
	\begin{minipage}[t]{0.23\linewidth}
	   \centering
	   \includegraphics[width=\textwidth]{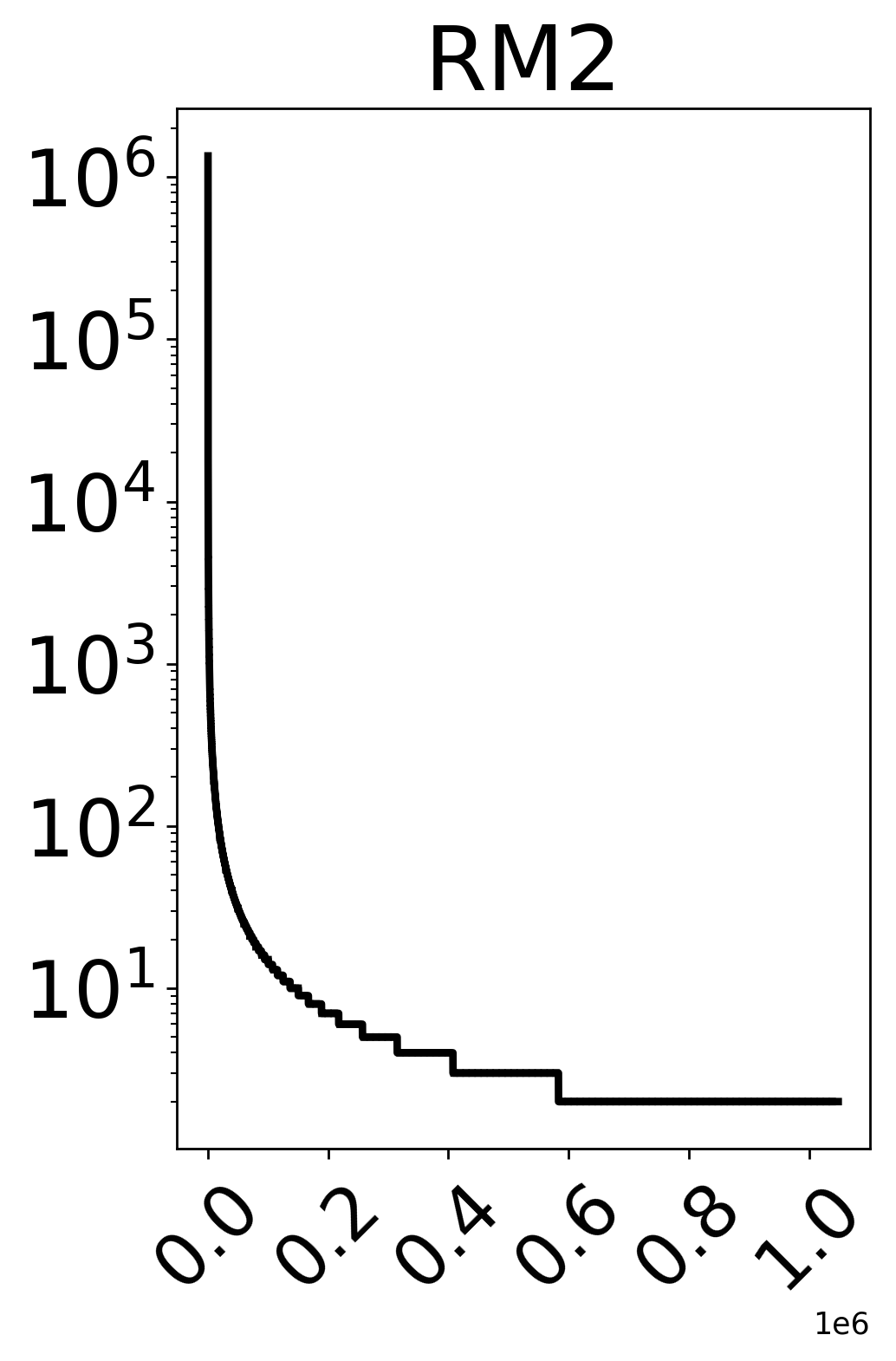}
	\end{minipage}
	\begin{minipage}[t]{0.23\linewidth}
	   \centering
	   \includegraphics[width=\textwidth]{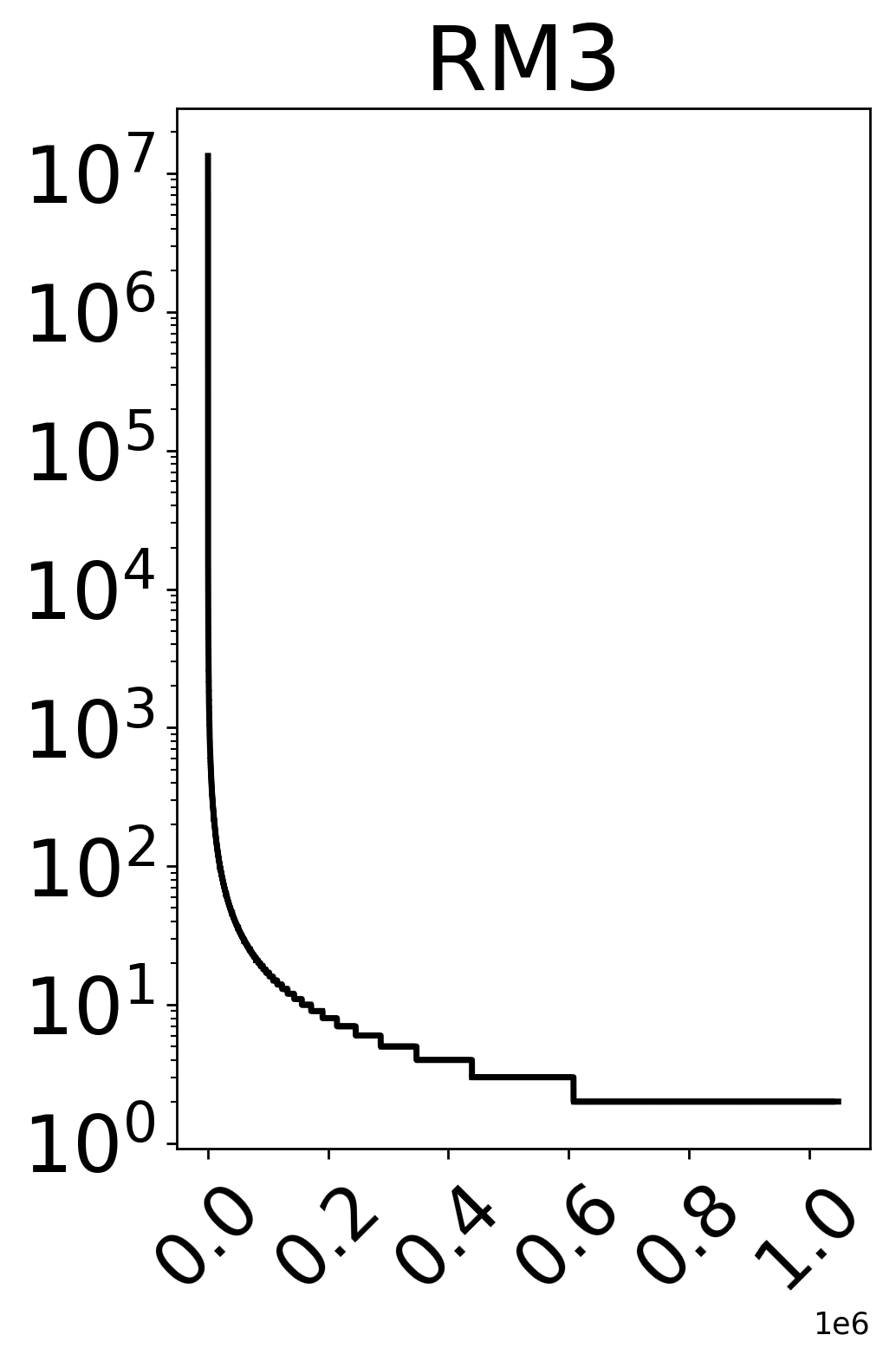}
	\end{minipage}
	\begin{minipage}[t]{0.23\linewidth}
	   \centering
	   \includegraphics[width=\textwidth]{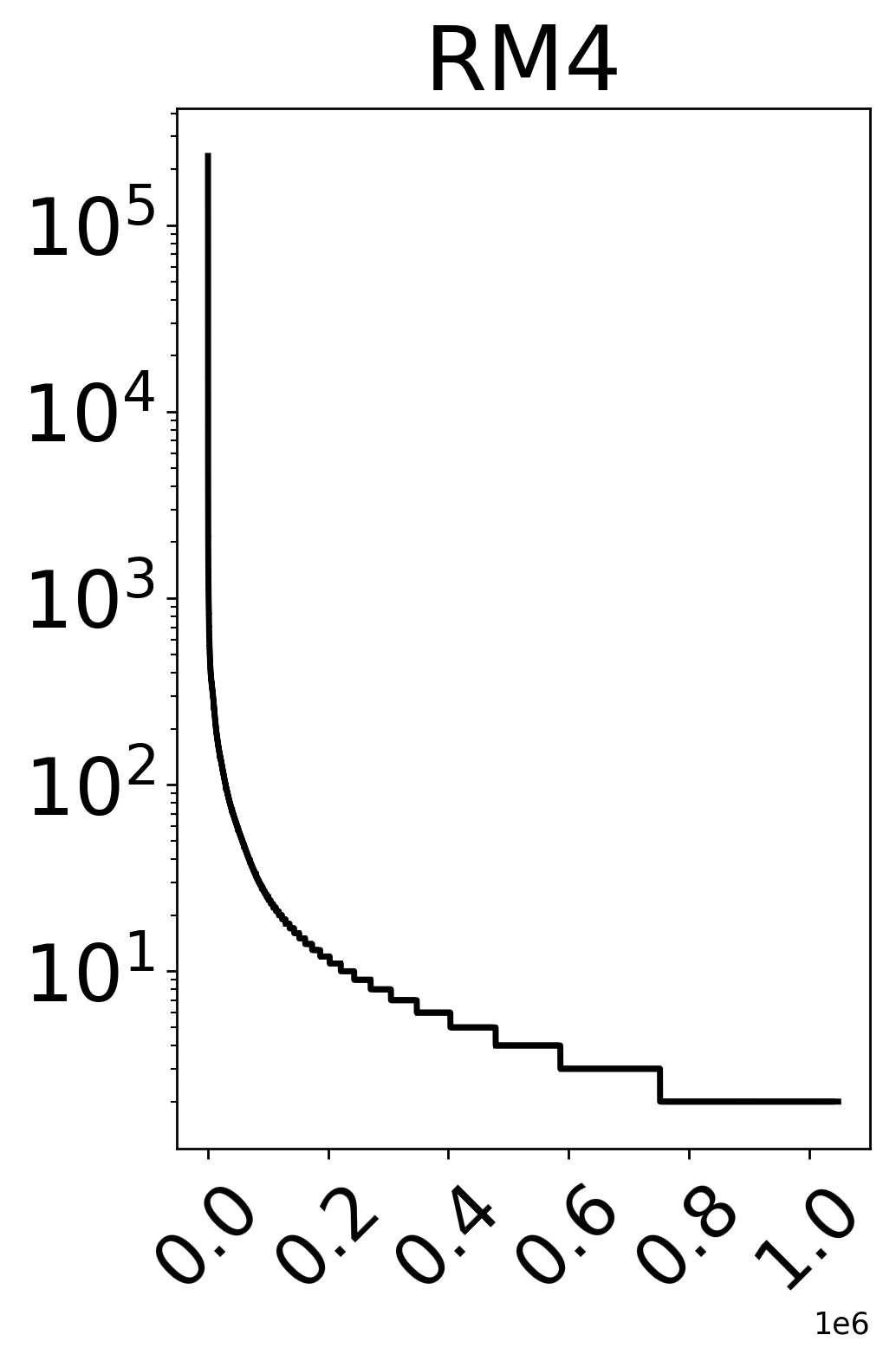}
	\end{minipage}
\caption{Access frequency to the largest embedding table during a single training epoch. This skewed access categorizes embeddings into `hot' and `cold.' The x-axis shows embedding indices in millions.}

\label{fig:access_skew}
\end{figure}

\begin{figure*}
\centering
    \includegraphics[width=0.85\textwidth]{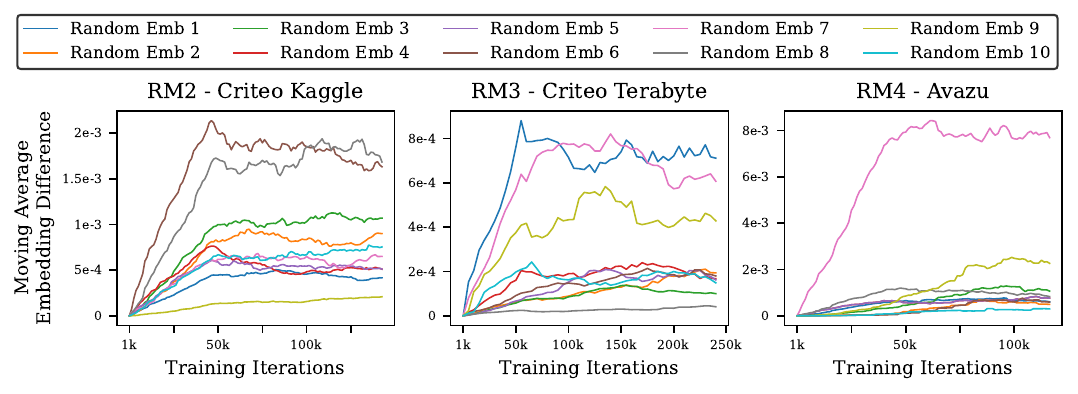}
	\caption{The temporal difference in values for \emph{ten randomly selected} `hot' embeddings for RM2 (Criteo Kaggle), RM3 (Criteo Terabyte), and RM4 (Avazu) recommendation models. As `hot' embeddings account for a significant fraction of accesses, they tend to saturate quickly -- in under 25\% of the training iterations. This experiment uses DLRM~\cite{dlrm} for the training process.}
    \label{fig:embeddingupdates}
\end{figure*}

\subsubsection{`Hot' Embeddings and Skewed Access Patterns}
We aim to capitalize on a fundamental aspect of recommendation models, where certain users and items demonstrate substantially higher popularity than others. This trend results in specific embeddings being accessed far more frequently than the rest, as depicted in Figure~\ref{fig:access_skew} across various real-world datasets. A small subset of 'hot' embeddings can typically receive over 100$\times$ more access than others. 

\subsubsection{Embedding Value Saturation} Given the dramatically higher access rates of `hot' embeddings compared to others, it is logical to expect them to converge faster during training. Therefore, our study focuses on identifying training phases where specific "hot" entries may exhibit stagnation. We systematically sampled entries from the hot embedding table across diverse datasets to explore this phenomenon and monitored their moving averages. 

The results, shown in Figure~\ref{fig:embeddingupdates}, offer insights into this behaviour. It becomes apparent that as training progresses, certain hot embedding entries plateau, showing minimal updates in magnitude. This suggests that after certain stages of training, any updates to these embeddings may not lead to significant performance enhancements in the model.

\subsection{Challenges: Identification of Stale Embeddings}

Low-cost identification of stale embeddings encounters two key challenges.

\subsubsection{Capturing Embedding Variations} The substantial memory requirements of embedding tables pose a significant obstacle in capturing temporal variations in embedding values. This challenge is twofold: first, we must isolate the ``hot" embeddings and then identify the stale embeddings within this subset. Let $\mathbf{E} \in \mathbb{R}^{m \times d}$ denote the embedding table, where $m$ represents the number of entries (rows), and $d$ denotes the embedding dimension. However, storing multiple temporal snapshots of the embedding table, denoted as $\mathbf{\hat{E}}$, is impeded by the sheer size of these tables (as we may need several snapshots), typically ranging from hundreds of gigabytes to terabytes (in commercial implementations).

\begin{table}[h!]
\captionof{table}{Memory Footprint of Embedding Tables}
    \resizebox{0.85\columnwidth}{!}{
    \begin{tabular}{|l|c|c|}
         \hline
         \multirow{2}{*}{\textbf{Model (Dataset)}} & \multirow{2}{*}{\textbf{Footprint}} & \textbf{\# Rows}  \\
         & &  (Biggest Table) \\
         \hline
         RM1 (Taobao Alibaba) & 0.3~GB & 4.1~Million\\
         RM2 (Criteo Kaggle) & 2~GB & 10.1~Million \\
         RM3 (Criteo Terabyte) & 63~GB & 11.9~Million\\
         RM4 (Avazu) & 0.55~GB & 6.6~Million\\
         \hline
    \end{tabular}}
    \label{tab:challenge1}
\end{table}

Thus, storing multiple snapshots in memory imposes substantial capacity and data movement overheads, rendering it impractical. Table~\ref{tab:challenge1} illustrates the memory footprint of the largest embedding tables across various models. For instance, the RM3 model exhibits an embedding footprint of 63GB. Notably, non-public commercial models commonly feature terabyte-sized embedding tables~\citep{baidu}, exacerbating the challenge further.

\begin{figure}[b]
	\centering
	\includegraphics[width=1\columnwidth]{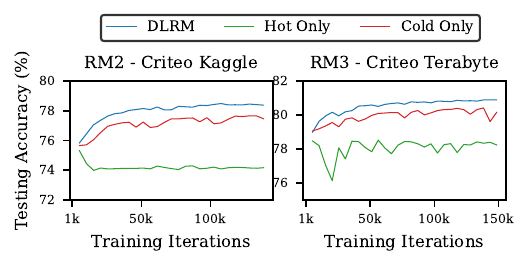}
	\captionof{figure}{Impact on testing accuracy when completely skipping cold or hot embedding updates compared to a baseline DLRM implementation. This representative analysis uses RM2 (Criteo Kaggle) and RM3 (Criteo Terabyte). Thus, we observe that a naive approach of skipping `hot' or `cold' embeddings can cause a significant accuracy loss of 4-6\%.}
	\label{fig:challenge2}
\end{figure}

\subsubsection{Determining Which Updates to Skip} 
Even if storing temporal snapshots of embedding entries is feasible, identifying embeddings that have undergone sufficient training poses a significant challenge. This challenge arises due to the uncertainty surrounding the optimal threshold $T$ for classifying embeddings as stale. A low value of $T$ may classify only a few embeddings as stale, maintaining baseline accuracy but potentially minimizing performance benefits from skipping stale embeddings. Therefore, an iterative assessment of each entry in $\mathbf{E}$ is necessary to determine an appropriate value of $T$ that balances baseline accuracy with performance benefits.

However, this iterative search incurs substantial computational overhead, potentially negating the performance gains from skipping stale embeddings. Alternatively, skipping `hot' or `cold' embeddings altogether could bypass these iterative overheads. Nevertheless, as illustrated in Figure~\ref{fig:challenge2}, simply skipping updates for `cold' or 'hot' embeddings leads to a significant reduction in accuracy, ranging from 4-6\% on datasets such as Criteo Kaggle and Criteo Terabyte.

%% file: body/slipstream.tex
\section{Design: The Slipstream Framework}
The Slipstream framework provides two significant performance advantages: (1) Eliminating compute and memory operations related to low-variability embeddings. (2) Enhancing CPU-GPU bandwidth utilization, reducing communication time overheads, and minimizing synchronization overhead. The \emph{key} notations used to describe the terms in the Slipstream framework design are summarized in Table~\ref{table:notations}.

\input{body/tables/notations}

\begin{figure*}
	\centering
	\includegraphics[width=1\textwidth]{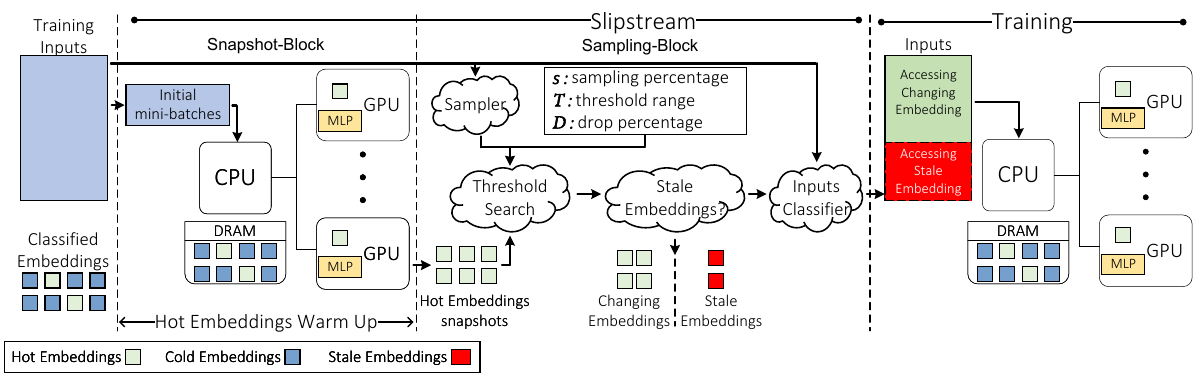}
	\caption{The flow for Slipstream. It consists of three phases. The first phase warms up the embeddings. This phase is similar to baseline training. The second phase samples inputs and converges on the threshold to determine the stale embeddings. The third phase drops inputs to stale embeddings and continues execution on only the varying embeddings.}
	\label{fig:slipstream}
 \vspace{-0.1in}
\end{figure*}

\subsection{Efficient Snapshots with `Hot' Embeddings}

Slipstream leverages the distinction between hot and cold embeddings, determined by their frequency of accesses, to create embedding snapshots. Let \(N\) denote the total number of snapshots. Notably, the hot embeddings, denoted by \(\mathbf{E}_{\text{hot}}\), are significantly smaller in size compared to the entire set of embedding tables \(\mathbf{E}\). For instance, in the terabyte dataset, \(|\mathbf{E}_{\text{hot}}|\) occupies only 500MB, yet caters to approximately 75\% of the total inputs. In contrast, the total size of the embedding tables is 63GB, more than 128 times larger than hot embeddings. Slipstream employs a module akin to the prior work FAE~\citep{fae} to identify the hot embeddings. We introduce the parameter \(\lambda\) to discern if an embedding entry is hot, defined as the number of accesses to an embedding entry to the total number of embedding accesses. For example, setting \(\lambda\) to 10$^{-5}$ would classify all embeddings with accesses \(\geq 1\) in 10000 as `hot' embeddings. This helps to identify hot embeddings (\(\mathbf{E}_{\text{hot}}\)) and  divide input training dataset $I$ into hot and cold input set as $I = I_{\text{hot}} + I_{\text{cold}}$. 

\begin{figure}[b]
	\centering
	\includegraphics[width=0.85\columnwidth]{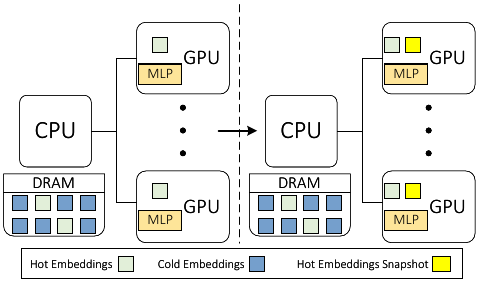}
	\caption{Creating snapshots of `hot' embeddings. Snapshots help Slipstream create a temporal profile of the embedding values. These snapshots now have a small footprint, presenting low memory capacity overheads within the GPU(s).}
	\label{fig:snapshot}
\end{figure}

\subsubsection{Warmup Period} Figure~\ref{fig:slipstream} shows the workflow of the Slipstream framework. The snapshot block is instantiated following a \emph{warmup} period, during which a comprehensive coverage of most embeddings is achieved with at least one access. The duration of the warmup period is determined empirically to ensure that a substantial portion of the inputs has been processed. Our experiments maintain a warmup period of 2K iterations (typically 1\%-2\% of the total iterations). The warmup period can be varied and is a hyperparameter to the Slipstream framework. Upon completion of the warmup period, Slipstream uses the \emph{low-footprint} temporal profiles in \(\hat{\mathbf{E}}_{\text{hot}}^n\) to inform subsequent training iterations.

\subsubsection{Embedding Snapshots} Thereafter, Slipstream utilizes \(N\) temporal instances, referred to as snapshots, denoted as \(\hat{\mathbf{E}}_{\text{hot}}^n\), for hot embedding table. Each snapshot captures a temporal profile of the embedding values at a specific point in time, where \(n\) ranges from 1 to \(N\). The size of each snapshot, \(|\hat{\mathbf{E}}_{\text{hot}}^n|\), is significantly smaller compared to \(|\mathbf{E}|\) and can be tailored to fit the available memory capacity of the GPU(s). Figure~\ref{fig:snapshot} shows the data placement after creating temporal snapshots of `hot' embeddings.

\subsection{Challenge: Identifying the Skip Threshold}

Slipstream aims to understand the right threshold (\(T\)) to skip the embedding entries that no longer need updating. We call these embeddings as stale embeddings. It can be done by comparing the respective `hot' embedding entries (\(\hat{\mathbf{E}}_{\text{hot}}\)) in snapshot \(n\) with an earlier snapshot \(n-1\) for all the inputs. We represent this as \(\Delta\hat{\mathbf{E}}_{\text{hot}} = \hat{\mathbf{E}}_{\text{hot}}^n - \hat{\mathbf{E}}_{\text{hot}}^{n-1}\). 

\begin{figure}[h!]
	\centering
	\includegraphics[width=0.85\columnwidth]{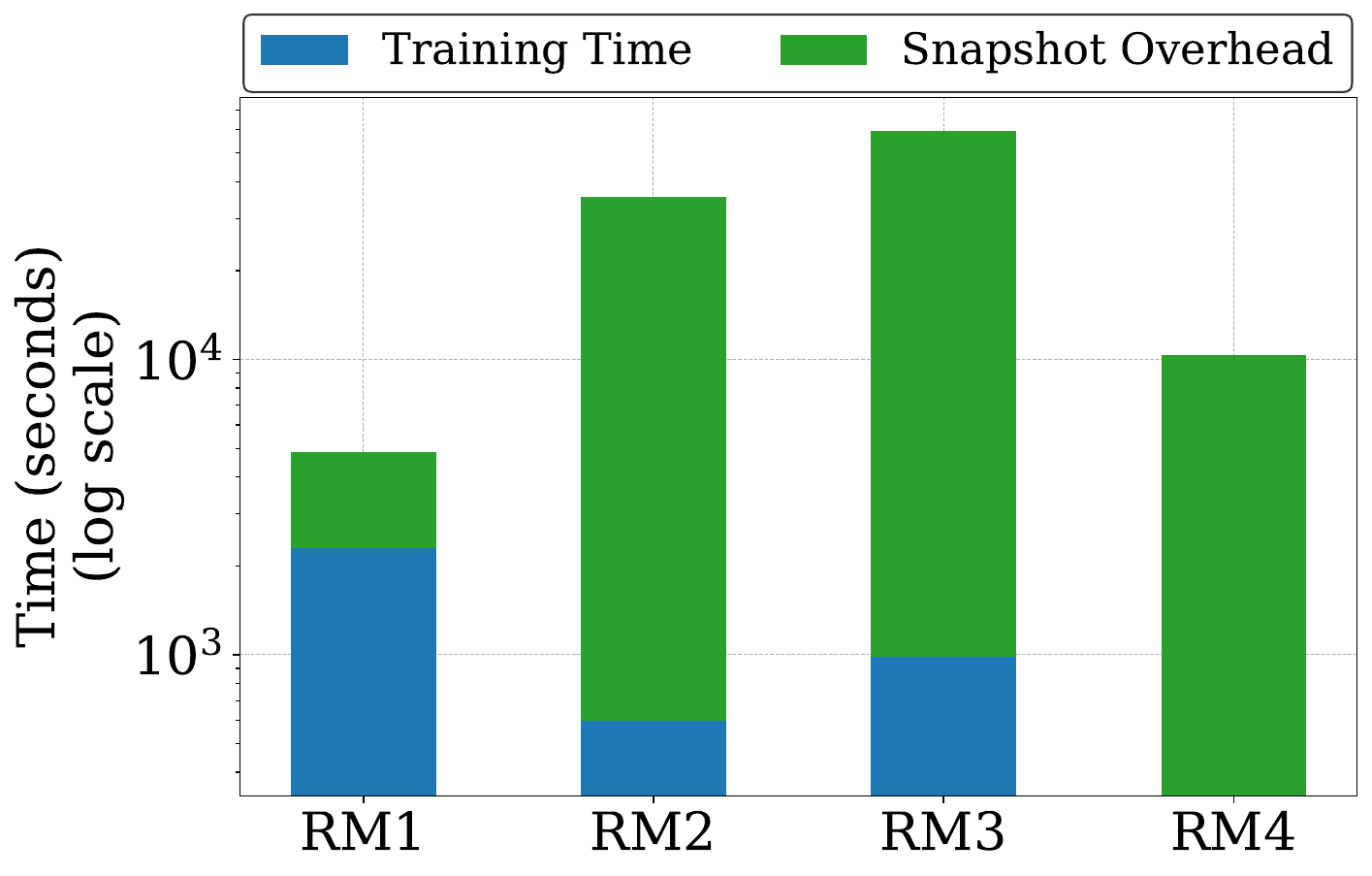}
	\caption{The overheads for creating temporal snapshots (log scale) of only `hot' embedding entries after each training minibatch. Comparing entries across different snapshot instances for different thresholds (\(T\)) presents 2$\times$ to 500$\times$ higher overheads than baseline training.}
	\label{fig:overheads_profiling}
\end{figure}

Figure~\ref{fig:overheads_profiling} shows how this process incurs a significant overhead on top of the baseline training time. This is because Slipstream would need to iterate over \emph{all} the `hot' embedding entries for different values of 
\(T\). For instance, the profiling overhead is $>$30$\times$ of the original training time for RM2 (Criteo Kaggle). For larger datasets such as RM3 (Criteo Terabyte), it tends to be $>$500$\times$.

\subsection{Solution: Sampling for Threshold Identification} 

\subsubsection{Insight} We leverage the insight that obtaining a precise threshold (\(T\)) may not be necessary. The Slipstream framework aims to converge on a (\(T\)) value with a certain confidence level to reduce the performance overheads in finding the appropriate threshold. Slipstream achieves this through input sampling. 
Instead of iterating over the entire embedding table across numerous inputs, Slipstream samples a subset of hot inputs $I_{\text{hot}}$. It then iterates over the embedding tables only for these sampled inputs.

\subsubsection{Input Sampling Block} The input sampling block randomly selects $s$ inputs from $I_{\text{hot}}$ inputs. Denoting the set of sampled inputs as $\mathcal{S}$, we have $|\mathcal{S}| = s$. The embeddings corresponding to these sampled inputs are then used for threshold-based comparison for different values of \(T\). For a given embedding threshold $T$, the inputs in $\mathcal{S}$ compare their indexed embedding vectors in each embedding snapshot.

We can represent the comparison of embedding entries between two snapshots, denoted as $\mathbf{\hat{e}}_{\text{hot}}^n$ and $\mathbf{\hat{e}}_{\text{hot}}^{n-1}$, using a threshold $T$. The comparison can be defined using Equation~\ref{eqn:delta}.

\begin{equation} \label{eqn:delta}
\delta(\mathbf{\hat{e}}_{\text{hot}}^n, \mathbf{\hat{e}}_{\text{hot}}^{n-1}, T) = 
\begin{cases}
    1, & \text{if } \|\mathbf{\hat{e}}_{\text{hot}}^n - \mathbf{\hat{e}}_{\text{hot}}^{n-1}\| > T \\
    0, & \text{otherwise}
\end{cases}
\end{equation}

Here, $\|\cdot\|$ represents a suitable distance metric (Slipstream uses the Euclidean distance), and $\delta(\cdot)$ indicates whether the embeddings have changed significantly (1) or not (0) based on the threshold $T$. 

\subsubsection{Dropping Inputs to Stale Embeddings} By iterating over the sampled inputs $\mathcal{S}$, Slipstream can now compute the \emph{fraction of total inputs} that can be dropped using the threshold $T$. To calculate this, let $D$ represent this proportion, given by Equation~\ref{eqn:drop}.

\begin{equation} \label{eqn:drop}
D = \frac{1}{|\mathcal{S}|}\sum_{s \in \mathcal{S}} \delta(s, T)
\end{equation}

We can also calculate the \emph{average number} of total hot inputs dropped. This is denoted as $\bar{D}$ and can be expressed as $\bar{D} = D \times I_{\text{hot}}$. Furthermore, we can calculate the input's standard deviation, denoted as $sd$, to capture the variability in the number of inputs dropped using Equation~\ref{eqn:dropsd}.

\begin{equation} \label{eqn:dropsd}
sd = \sqrt{\frac{1}{|\mathcal{S}|}\sum_{s \in \mathcal{S}} \left(\delta(s, T) - D\right)^2}
\end{equation}

\subsubsection{Confidence in Estimating Dropped Inputs} We use the "Student's t-interval" to establish a confidence interval (CI) in our estimation. Let $\bar{\emph{D}}$ follow a t-distribution. The 100$\times$(1-$\alpha$) CI for $\bar{\emph{D}}$ is given by Equation~\ref{eqn:ciformula}.

\begin{equation}  \label{eqn:ciformula}
CI_{100\times(1-\alpha)} = \bar{D} \pm t_{\frac{\alpha}{2}} \times \sqrt{\left(\frac{I-|\mathcal{S}|}{I}\right) \times \left(\frac{sd^2}{m}\right)}
\end{equation}

Here, $CI_{100\times(1-\alpha)}$ represents the CI for the estimated fraction of dropped inputs, $\bar{D}$. The value of $t_{\frac{\alpha}{2}}$ corresponds to the critical value of the t-distribution with $\frac{\alpha}{2}$ degrees of freedom. $m$ indicates the number of input samples. For a 99.9\% confidence level, we approximate $t_{\frac{\alpha}{2}}$ as 3.340.

\subsubsection{Rapidly Finding Threshold} 
\begin{figure}[h!]
	\centering
	\includegraphics[width=1\columnwidth]{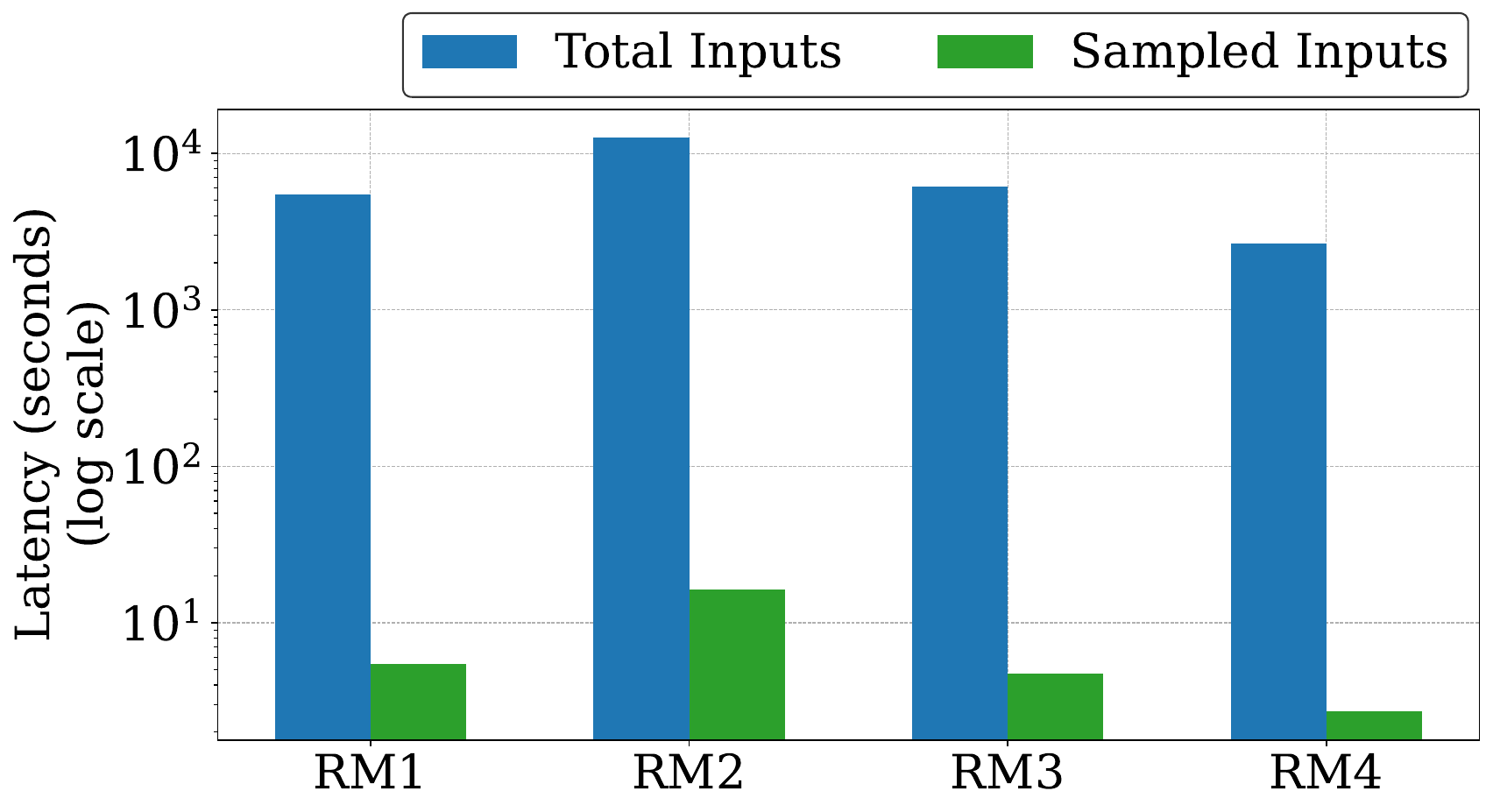}
	\caption{The latency savings due to sampling for four real-world datasets. We observe at least a 1000$\times$ lower latency as the sampling block selects 0.1\% of hot inputs ($I_{\text{hot}}$).}
	\label{fig:latency saving}
    \vspace{-0.1in}
\end{figure}

Slipstream uses a binary search algorithm to determine an appropriate threshold $T$. The search process iteratively tests different threshold values, utilizing only the sampled inputs $\mathcal{S}$. Slipstream identifies the optimal threshold by comparing the number of dropped inputs at each threshold, resulting in substantial computational savings. Figure~\ref{fig:latency saving} shows the latency savings achieved by the input sampling block with 0.1\% input samples compared to a naive approach that examines all inputs. The input sampling block reduces overheads compared to naively accessing \emph{all} the `hot' embeddings by nearly 1000$\times$.

\subsection{Input Classifier Block}
The input classifier block separates inputs that access varying embeddings from inputs that access stale embeddings. The corresponding embedding vectors from consecutive snapshots are compared for each input $i \in I_{\text{hot}}$. Let us assume that Slipstream takes $N$ snapshots. Let $\hat{\mathbf{E}}_{\text{hot}}^n (i)$ denote the embedding matrix associated with input $i$ for nth snapshot. If the Euclidean distance between the embedding vectors at consecutive snapshots $n$ and $n-1$ is greater than threshold $T$, the corresponding embedding entry $\mathbf{e}_{\text{hot}} (i)$ is marked as varying. This can be expressed using Equation~\ref{eqn:snapshotdetail}.

\begin{equation} \label{eqn:snapshotdetail}
\left\| \mathbf{E}_{\text{hot}}^n (i) - \mathbf{E}_{\text{hot}}^{n-1} (i) \right\|_2 > T \Rightarrow \mathbf{E}_{\text{hot}}(i) \text{ is varying}  
\end{equation}

Based on the marked hot embeddings, hot inputs $I_{\text{hot}}$ are classified into two subsets given a hyperparameter $\alpha$ that defines the number of stale features accessed by input $i$ to be skipped. First, inputs that access embeddings that are marked as varying ($I_{\text{hot}}^{\text{vary}}$). $I_{\text{hot}}^{\text{vary}}$ can be defined using Equation~\ref{eqn:vary}.
 \begin{equation} \label{eqn:vary}
 I_{\text{hot}}^{\text{vary}} (\alpha) = \{i (\alpha) \in I_{\text{hot}} \mid \mathbf{E}_{\text{hot}}(i) \text{ is varying}\}
 \end{equation}

Second, inputs that access embeddings that are marked as stale ($I_{\text{hot}}^{\text{stale}}$). $I_{\text{hot}}^{\text{stale}}$ can be defined using Equation~\ref{eqn:nonvary}.
\begin{equation} \label{eqn:nonvary}
I_{\text{hot}}^{\text{stale}} (\alpha) = \{i (\alpha) \in I_{\text{hot}} \mid \mathbf{E}_{\text{hot}}(i) \text{ is stale}\}
 \end{equation}

 \begin{equation} \label{eqn:classify}
I_{\text{hot}} = I_{\text{hot}}^{\text{vary}} + \bcancel{I_{\text{hot}}^{\text{stale}}}
 \end{equation}

The input classifier block incurs a one-time computational overhead by performing the classification process in a single pass. Only the inputs in $I_{\text{hot}}^{\text{vary}}$ are utilized during the remaining training process. This is because they access embedding entries that exhibit variations.

\subsection{Accuracy Recovery using Feature Normalization}

The Slipstream framework has a certain confidence level in skipping stale embeddings. Consequently, it's conceivable that the framework might erroneously categorize a small fraction of non-stale embeddings as stale. This could lead to a marginal reduction in accuracy compared to the baseline that processes all inputs, primarily affecting metrics like the Area Under Curve (AUC). In DLRM-based training, AUC serves as a determinant for training completion, and any decline in AUC implies a loss in accuracy. Even a minor decrease in AUC holds significance in recommendation tasks~\citep{din, baidu}.

Slipstream employs feature normalization to counteract this. This ensures that all dense and sparse vectors have zero mean and unit variance. This approach stabilizes feature distribution and improves stale embedding detection. This also hastens convergence and modestly regularizes embedding vectors to reach baseline accuracy. Slipstream incorporates a LayerNorm layer into its framework to implement this.

\subsection{Tying it Together}
Algorithm~\ref{alg:slipstream_algo} describes the workflow of Slipstream. The algorithm begins by initializing various parameters, including the embedding table $E$, the hot embedding table $E_{\text{hot}}$, thresholds for hotness ($\lambda$), and staleness ($\theta$). Initially, Slipstream identifies the access pattern to embedding entries. In this stage, called the preprocessing phase, hot entries are moved to $E_{\text{hot}}$, and a model warm-up phase is executed. This phase samples a small percentage of hot inputs to train the model. During the warm-up phase, we take $N$ snapshots of hot embedding entries to track the variability of frequently accessed entries. A distance metric measures the change $s$, and the algorithm searches each `hot' embedding snapshot for an optimal staleness threshold $\theta$. Inputs are classified and selected based on their effect on non-stale `hot' entries. This process continues until a stopping condition is met. Overall, Slipstream improves training efficiency by selectively updating embeddings and prioritizing non-stale hot entries while minimizing unnecessary computations and data transfers.

\input{body/tables/ss_algo}

%% file: body/tables/notations.tex
\begin{table}[h!]
\centering
\caption{Table of Notations of \emph{Key} Terms}
\resizebox{0.9\columnwidth}{!}{
\begin{tabular}{| c | l |}
 \hline
 \textbf{Notation} & \textbf{Description} \\
 \hline
 $\mathbf{E}$ & All embedding tables \\
 $\mathbf{E}_{\text{hot}}$ & Hot embedding table \\
 $\hat{\mathbf{E}}_{\text{hot}}^n$ & nth snapshot of hot embedding table\\
 $\mathbf{\hat{e}}_{\text{hot}}^n (i)$ & ith entry of nth snapshot of hot embedding table \\
$N$ &  Total number of snapshots \\
$\lambda$ & Ratio of accesses for an embedding entry to be hot\\
 $T$ & Threshold to determine stale embeddings \\
 $I$ & Training inputs dataset \\
 $I_{\text{hot}}$ & Hot inputs set \\
 $I_{\text{cold}}$ & Cold inputs set \\
 $s$ & Sampling percentage from $I_{\text{hot}}$\\
$\mathcal{S}$ & Sampled input set from $I_{\text{hot}}$\\
 \textbf{\emph{$D$}} & Input drop-percentage (\%) from $\mathcal{S}$ \\
 \textbf{\emph{$\bar{D}$}} & Average input drop-percentage from $I_{\text{hot}}$\\
 $\alpha$ & Number of stale features for an input to be skipped \\
 \textbf{\emph{$sd$}} & Standard Deviation \\
 \textbf{\emph{$CI$}} & Confidence Interval \\

 \hline
\end{tabular}
}
\label{table:notations}
\end{table}

%% file: body/tables/ss_algo.tex
\begin{algorithm}
\caption{Slipstream Workflow}
\label{alg:slipstream_algo}
\footnotesize

\begin{algorithmic}[1] 

\STATE \textbf{Embedding Table}: $\mathbf{E}$$ \quad \# $\text{E(i) is ith embedding entry}

\STATE \textbf{Counter}: $\mathbf{C_i}$$
\quad \# $\text{Access counter for ith embedding entry $\text{E(i)}$}

\STATE \textbf{Hot Embedding Table}: $\mathbf{E}_{\text{hot}}$

\STATE \textbf{Hotness threshold}: $\lambda$ \quad \# Access ratio for an entry to be hot or not

\STATE \textbf{Staleness threshold}: $T$ \quad \# Threshold for hot entry to be stale or not

\STATE \textbf{nth Snapshot of hot embedding entry}: $\hat{\mathbf{E}}_{\text{hot}}^{n}$ \quad

\STATE \textbf{Input training dataset}: $I$ \quad

\STATE \textbf{Training minibatch}: mbs \quad

\STATE \textbf{$Preprocessing$ $phase$}:

\WHILE {$i$ $\in$ $I_{\text{sample}}$ }
    \STATE $C_i$ $\gets$ $E(i)$

    \IF [\# Access ratio of entry is greater than $\lambda$ ]{$\mathbf{C_i}$ $\ge$ $\lambda$ }
    \STATE  $\mathbf{E_{\text{hot}}}$ $\gets$ $E(i)$ \# \textbf{Classify embedding as hot}
    \ENDIF
    
\ENDWHILE

\STATE \textbf{$Warm$ $up$ $phase$}:

\WHILE[\# Iterating through initial minibatches]{mbs $\mathbf{\in}$ 1-2 \% of batches }
\STATE \textbf{Train}(mbs)
\STATE $\hat{\mathbf{E}}_{\text{hot}}^{n} \gets \text{Snapshot}$

\ENDWHILE

\STATE \textbf{$Slipstream$ $phase$}:

\STATE $i = 0$:

\WHILE[Iterating through Snapshots]{$ i \le N $ }
\STATE \(\Delta\hat{\mathbf{E}}_{\text{hot}} (i) = \hat{\mathbf{E}}_{\text{hot}}^i - \hat{\mathbf{E}}_{\text{hot}}^{i-1}\)

\IF [\# Greater than threshold $T$] {$\Delta\hat{\mathbf{E}}_{\text{hot}} (i) > T$}

\STATE $\mathbf{E}_{\text{hot}}(i) \text{ is varying}$  
\ENDIF

\STATE \textbf{Increment($i$) }
\STATE \textbf{Train($S$) } \COMMENT{ Train model on sampled hot inputs}
\STATE \textbf{ $\mathbf{E}_{\text{hot}}(i) 
 \gets \delta(\mathbf{\hat{e}}_{\text{hot}}^n, \mathbf{\hat{e}}_{\text{hot}}^{n-1}, T) $} \COMMENT{ Equation ~\ref{eqn:delta}}

\ENDWHILE

\STATE $\textbf{$Calculate$ I}_{\text{hot}}^{\text{stale}} (\alpha) \COMMENT{ Equation ~\ref{eqn:nonvary}}$

\STATE $\textbf{Train I}_{\text{hot}}^{\text{vary}} (\alpha) \COMMENT{ Equation ~\ref{eqn:classify}}$

\end{algorithmic}
\end{algorithm}

%% file: body/evaluation.tex
\begin{table*}[h!]
\centering
\caption{The recommendation model architectures, DLRM~\citep{dlrm} and TBSM~\citep{tbsm}, and datasets evaluated for Slipstream. The table provides the corresponding dataset, input dense and sparse feature size, the top and bottom MLP network architecture, and embedding table details.}
\input{body/tables/models}
\label{tab:models}
\end{table*}

\begin{figure*}[h!]
  \centering
  \subfloat[RM1 - Taobao Alibaba]{
	\begin{minipage}[t]{0.245\textwidth}
         \centering
	   \includegraphics[width=\textwidth]{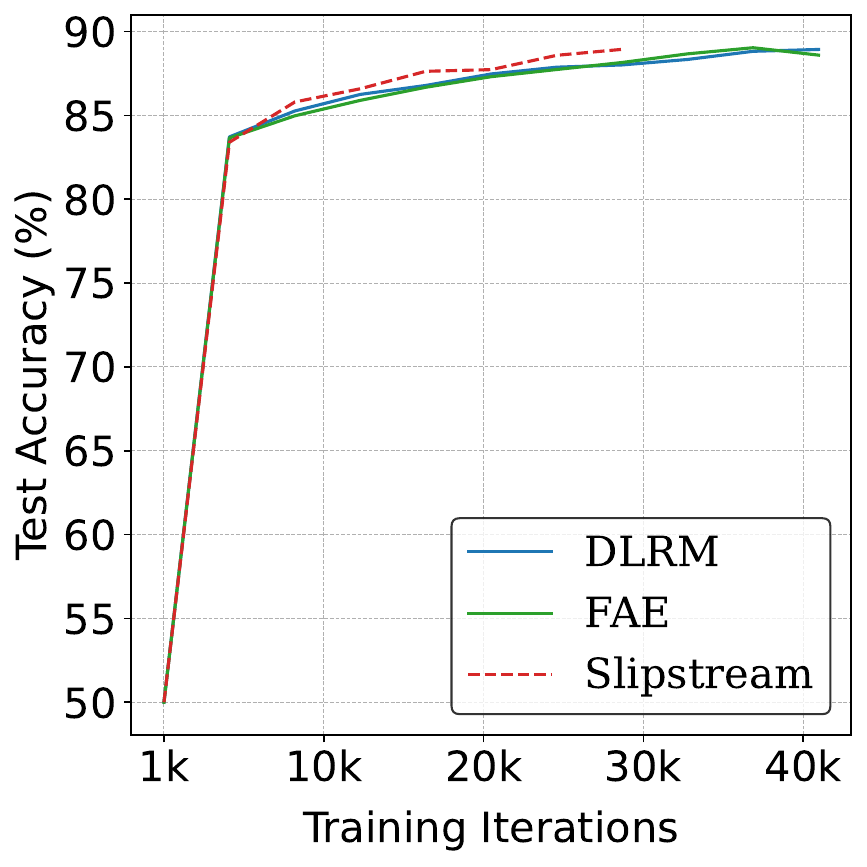}
	\end{minipage}}	
  \subfloat[RM2- Criteo Kaggle]{
	\begin{minipage}[t]{0.245\textwidth}
	   \centering
	   \includegraphics[width=\textwidth]{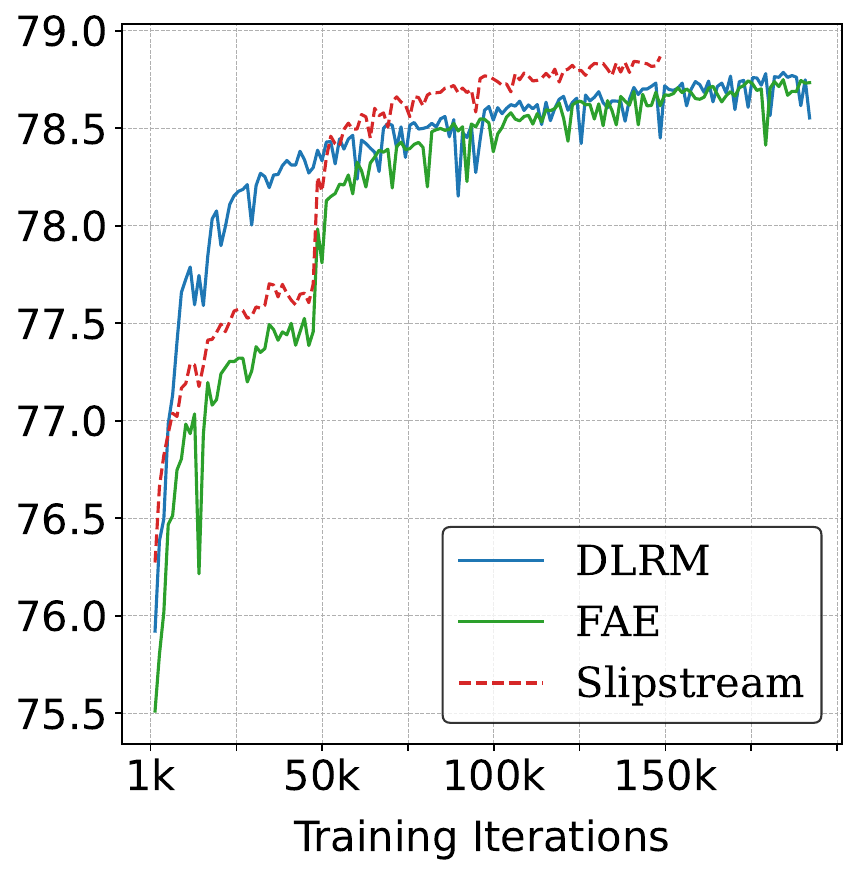}
	\end{minipage}}
  \subfloat[RM3 - Criteo Terabyte]{
	\begin{minipage}[t]{0.245\textwidth}
	   \centering
	   \includegraphics[width=\textwidth]{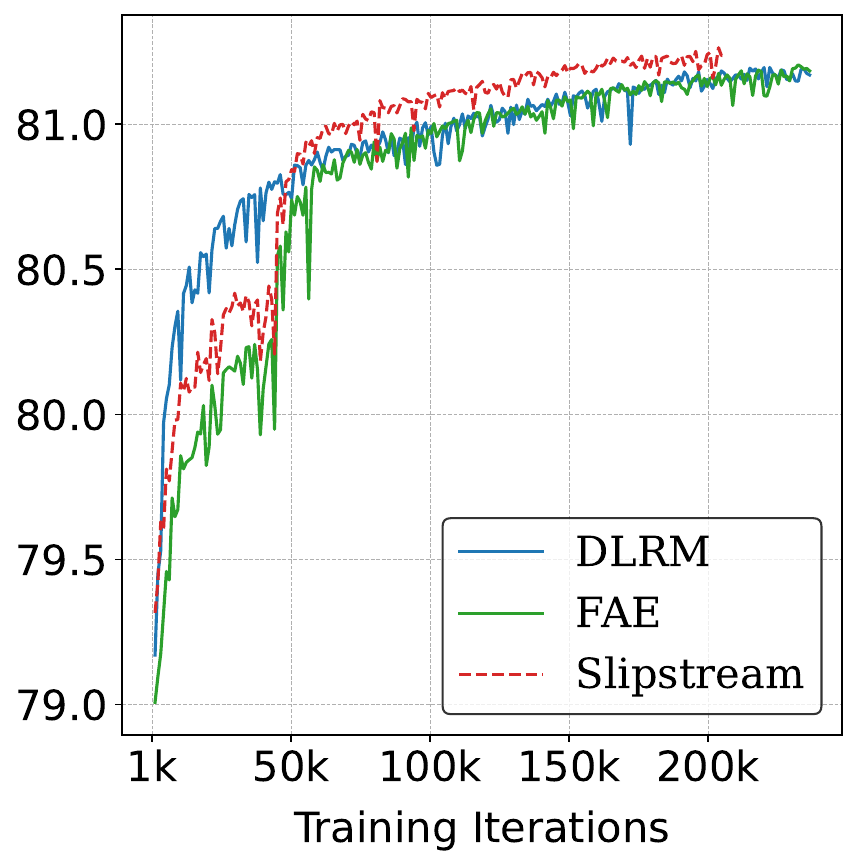}
	\end{minipage}}
  \subfloat[RM4 - Avazu]{
	\begin{minipage}[t]{0.245\textwidth}
	   \centering
	   \includegraphics[width=\textwidth]{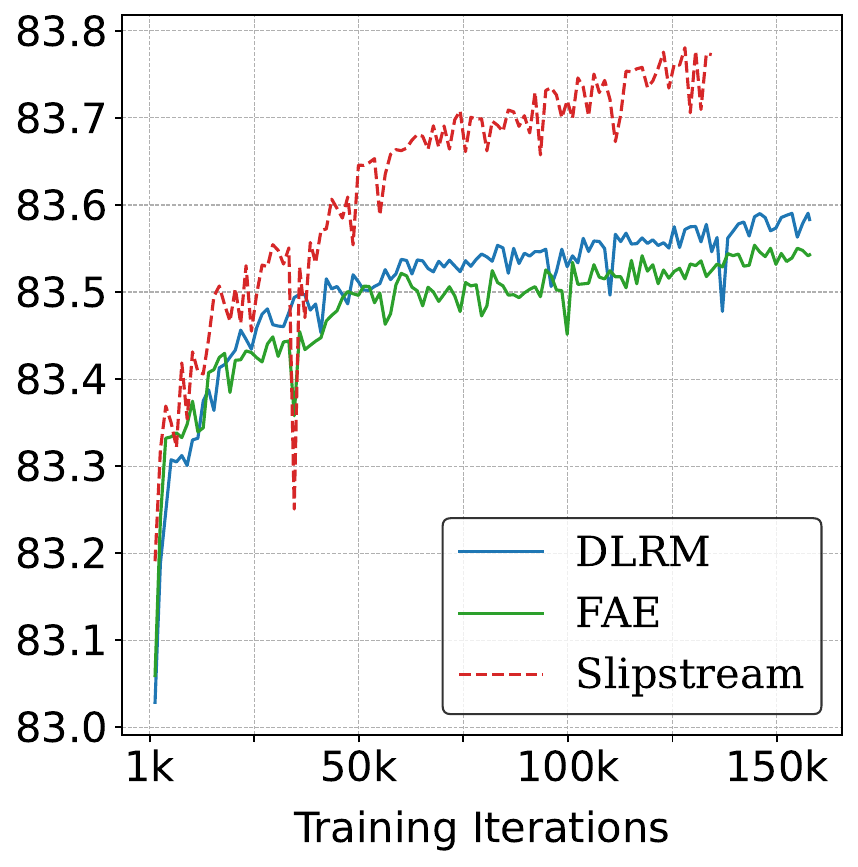}
	\end{minipage}}
\caption{Testing Accuracy across four real-world datasets and commonly used recommender models. The testing accuracy is measured across training iterations. It is noteworthy that Slipstream consistently improves the accuracy curve compared to DLRM and FAE baselines. Overall, Slipstream incurs an \textbf{average accuracy gain of 6\%}.}
\vspace{-0.1in}
\label{fig:accuracyresults}
\end{figure*}

\section{Evaluation}
\label{sec:evaluation}
\subsection{Experimental Setup and Models}
\label{subsec:models}

To evaluate the performance benefits of Slipstream, we used four different-sized recommendation models with four publicly available real-world datasets. We employ four well-established deep-learning recommendation models to evaluate Slipstream. These models (RM1, RM2, RM3, and RM4) cover a wide variety and size of recommendation models with dense features ranging from 1 to 13 while sparse features range from 3 to 26. Similarly, these models present a range of neural networks with varying depth and width. Table~\ref{tab:models} shows each model's detailed architecture and features. Open-sourced deep learning recommendation model (DLRM)~\citep{dlrm} and time-based sequence model (TBSM)~\citep{tbsm} are used to train these recommendation models. 

The hyperparameter $\lambda$ determines whether an embedding entry is popular. This hyperparameter is used in the FAE framework ~\citep{fae}. It dictates the size of the hot embedding entry on the GPU. Slipstream chooses $\lambda$ equals to 10$^{-6}$ for RM1 and 10$^{-7}$ for RM2, RM3, and RM4 to balance the hot embedding size on the GPU and the percentage of hot inputs. For instance, a $\lambda$ value of 10$^{-6}$ dictates that embeddings with more than 1 in a million accesses are deemed hot or popular.

\subsubsection{Datasets}

Slipstream is evaluated using four publicly available real-world datasets to train recommendation models -- namely Taobao~\citep{alibaba}, Criteo Kaggle~\citep{criteokaggle}, Criteo Terabyte~\citep{criteoterabyte}, and Avazu~\citep{avazu}. The Taobao dataset contains user behaviours from the Taobao e-commerce platform, the Criteo Kaggle dataset involves predicting click-through rates for display ads, the Criteo Terabyte dataset is a large-scale click log dataset used as an MLPerf benchmark, and the Avazu dataset is derived from a click-through rate prediction competition.

\subsubsection{Baselines}

We compare against four state-of-the-art baselines: (1) XDL: An Industrial Deep Learning Framework for High-dimensional Sparse Data~\citep{xdl}, (2) An open-source implementation of DLRM and TBSM~\citep{dlrm, tbsm}, (3) FAE: A software framework that exploits input popularity~\citep{fae}, and (4) Hotline: A hardware accelerator that pipelines popular and non-popular inputs~\citep{hotline}. All baselines are executed in CPU-GPU hybrid mode, with embeddings executed on the CPU and neural networks on the GPU(s).

\subsubsection{Hardware and Software Setup}

The DLRM and TBSM code is configured using Pytorch-1.8.1 ~\citep{pytorch} and executed using Python-3.8. NCCL is used ~\citep{nccl} for gather, scatter, and all-reduce collective calls for GPU-GPU communication via the backend NVLink~\citep{nvlink}. XDL-1.0~\citep{xdl} is used with Tensorflow-1.2~\cite{tensorflow} as the computation backend.

Table~\ref{tab:systemconfig} shows the configuration of our evaluation hardware setup. The GPUs are connected using the high-speed NVLink-2.0 interconnect. Every GPU communicates with the rest of the system via a 16x PCIe Gen3 bus. This system is based on the state-of-the-art FAE setup ~\citep{fae}.

\input{body/tables/server}

\begin{figure*}[h!]
	\centering
	\includegraphics[width=1\textwidth]{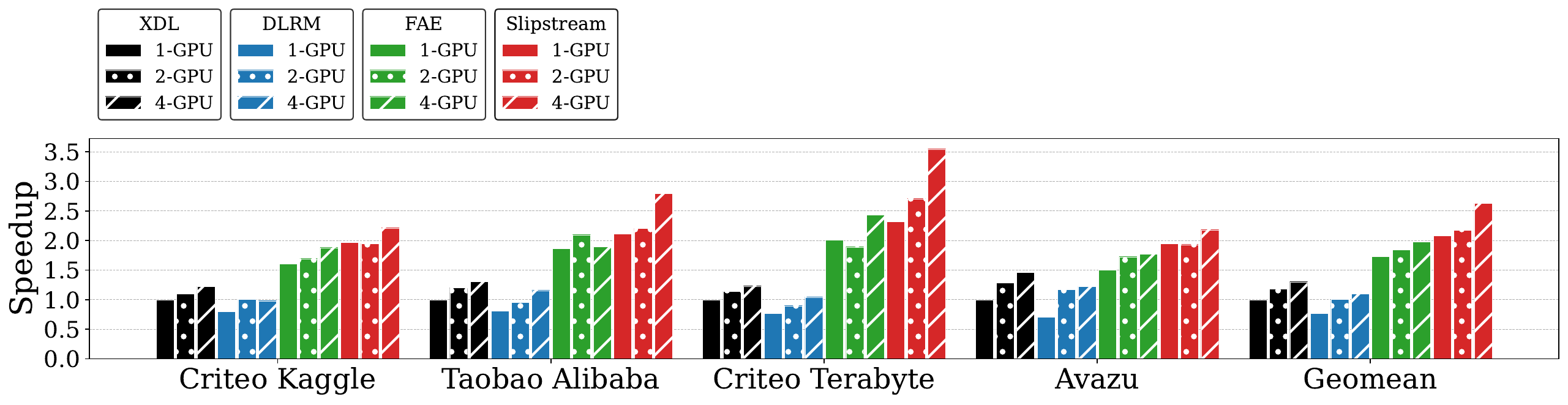}
	\caption{The performance of Slipstream compared to the prior state-of-the-art for Criteo Kaggle, Taobao Alibaba, Criteo Terabyte, and Avazu datasets. All values are normalized to XDL 1-GPU.  On average, Slipstream provides 2.5$\times$ training time speedup compared to an aggressive XDL baseline. It also consistently maintains a high speedup across all datasets and models.}
	\label{fig:performance}
\end{figure*}

\begin{figure*}[t]
	\centering
	\includegraphics[width=1\textwidth]{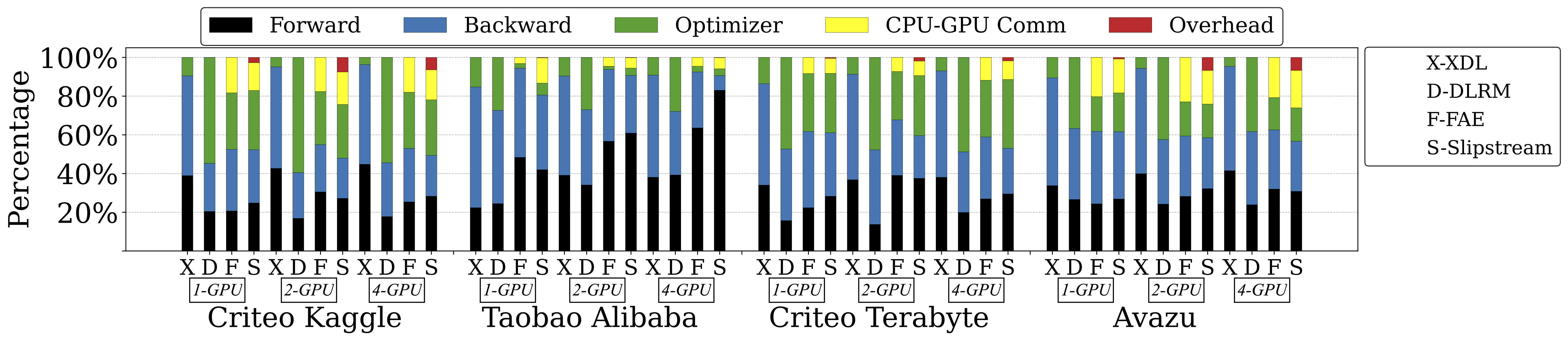}
	\caption{The performance breakdown of Slipstream compared to the prior state of the art prior work XDL, FAE, and DLRM for Criteo Kaggle, Taobao Alibaba, Criteo Terabyte, and Avazu datasets. We observe that Slipstream incurs a negligible overhead of a maximum of 5 percent of the total execution time for datasets such as Kaggle and Avazu.}
	\label{fig:perf_breakdown}
\end{figure*}

\subsection{Results and Analysis}
\subsubsection{Accuracy}
Figure~\ref{fig:accuracyresults} illustrates the testing accuracy results of Slipstream compared to the DLRM and FAE baselines. The baselines achieve an accuracy of 78.5\%, 89\%, 81\%, and 83.5\% for Criteo Kaggle, Taobao Alibaba, Criteo Terabyte, and Avazu datasets, respectively. Slipstream achieves higher accuracy than the baselines across all datasets, indicating its effectiveness in maintaining competitive performance. It also suggests that skipping inputs for saturated embeddings does not compromise accuracy. Slipstream exhibits an initial accuracy improvement for the Criteo datasets due to its differentiated treatment of popular and non-popular inputs. Thereafter, it exceeds the baseline accuracy curve and maintains a higher training accuracy.

\subsubsection{Evaluation Metrics}
The detailed numeric evaluation metrics -- Area Under Curve (AUC), Testing Accuracy and Binary Cross Entropy (BCE) loss -- are summarized in Table~\ref{tab:evalmetrics}. We observe that Slipstream largely exceeds these metrics compared to the baseline DLRM.

\input{body/tables/metric}

\subsubsection{Performance Improvement}
Slipstream achieves performance benefits by selectively skipping non-contributing inputs during training. Figure~\ref{fig:performance} shows that Slipstream consistently provides a 2.5$\times$ training time speedup compared to an aggressive XDL baseline across datasets and recommender models. Experimental results on 4-GPUs using weak scaling demonstrate that Slipstream's speedup improves as the number of GPUs increases. However, for the Taobao Alibaba dataset, which relies on sequence-based inputs, the benefits of Slipstream compared to FAE are less pronounced due to additional overhead. Overall, Slipstream reduces bandwidth overheads for collective communications, such as scatter and gather operations, improving training efficiency. 

\subsubsection{Performance Breakdown}
Figure~\ref{fig:perf_breakdown} shows the performance breakdown of Slipstream compared to the baseline. On average, Slipstream incurs a negligible overhead of a maximum of 5 percent of the total execution time for datasets such as Kaggle and Avazu. Overall, the Slipstream overhead remains within a tight bound. Therefore, the larger the model becomes, the ratio of Slipstream overhead to the total performance reduces. This is true for the case of the Alibaba dataset as well. Although the model is not considered large, the performance is high due to its sequential nature.

\subsubsection{Scaling mini-batch size}
\begin{figure}[h!]
	\centering
	\includegraphics[width=\columnwidth]{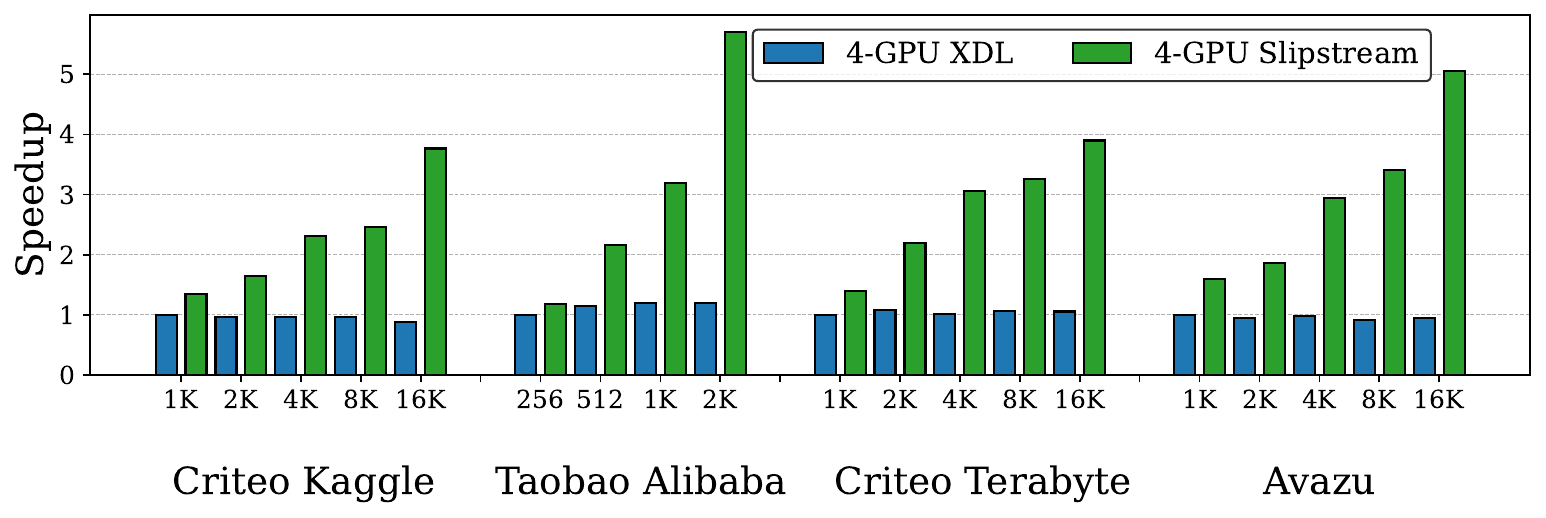}
	\caption{Mini-batch size scaling for 4-GPU execution. Slipstream showcases a minimum speedup of 3.8$\times$ considering all models and datasets as we increase the mini-batch size.}
	\label{fig:mbs_scaling}
\end{figure}

Figure~\ref{fig:mbs_scaling} illustrates the Slipstream speedup relative to the 4-GPU XDL baseline across the Kaggle, Taobao Alibaba, Criteo Terabyte, and Avazu datasets. The results indicate that time-based models like Taobao Alibaba exhibit better scalability, while models like Criteo Terabyte do not scale proportionally. Furthermore, in the case of Taobao Alibaba, it uses a smaller dataset that also features several embeddings considered hot. Thus, a significant fraction of these embeddings can be placed on the GPU, enhancing training. Thus, on average, Taobao Alibaba has multiple embedding table accesses and operations for each input, enabling these embeddings to train quickly and become stale. This enables Taobao Alibaba to achieve much higher speedup gains from Slipstream.

Slipstream's scalability stems from its reduction in communication overhead. Transitioning from a single GPU to multiple GPUs (4 GPUs), the collective communication overheads, such as all-to-all or one-to-all communications, outweigh the benefits of weak scaling. Additionally, larger mini-batches require higher bandwidth for data transfer, affecting the CPU-GPU communication links. 

\subsubsection{Comparison with Hardware Accelerator}

Prior work has proposed employing a hardware accelerator, Hotline~\citep{hotline}, to pipeline the cold embeddings while inputs to hot embeddings are processed on the GPU. The Slipstream software framework is orthogonal to this approach. We implemented Slipstream on the top of Hotline for a 1-GPU system. Table~\ref{tab:hotline} shows that Slipstream provides 17.5\% speedup across publicly available datasets and models. Furthermore, Slipstream is a software framework that can be readily deployed, Hotline is a specialized hardware accelerator that needs to be designed and incorporated into the system.
\begin{table}[h!]
\caption{Enabling Slipstream on Hotline (1-GPU)}
    \resizebox{0.4\columnwidth}{!}{
    \begin{tabular}{|c|c|}
         \hline
         \textbf{Model} & \textbf{Speedup on the}\\
         \textbf  & \textbf{top of Hotline}\\
         \hline
         RM1 & 18\%\\
         RM2 & 14\% \\
         RM3 & 19\% \\
         RM4 & 18\% \\
         Geomean & 17\%\\
         \hline
    \end{tabular}}
    \label{tab:hotline}
\end{table}

\subsubsection{Ablation Study: Slipstream Accuracy Recovery}
\label{subsubsec:slipstream_ln}
To assess the influence of misidentified stale embeddings by Slipstream and the efficacy of feature normalization on convergence and accuracy, we conducted experiments where we omitted the LayerNorm layer from Slipstream while keeping the rest of the architecture unchanged. The impact on the overall evaluation metrics with and without LayerNorm is summarized in Table~\ref{tab:layer_norm}. The removal of LayerNorm reduced AUC and accuracy, slightly below the baseline established by DLRM. However, the incorporation of LayerNorm not only matched but surpassed baseline accuracy for specific models and datasets, such as RM4. This led to accelerated convergence and enhanced detection of stale embeddings through normalization.

\input{body/tables/slipstream_ln}

\subsubsection{Ablation Study: Hyperparameter $\alpha$}
\label{subsubsec:alpha}

\input{body/tables/alpha_drop_percentage}

Slipstream skips inputs that access stale embeddings, with the number of skipped inputs influenced by a parameter called $\alpha$.  $\alpha$ varies from 0 to the total number of sparse features. This parameter significantly affects the drop percentage ($D$) and accuracy. Table~\ref{tab:alpha2} shows ablation studies for varying $\alpha$ values impact drop percentage ($D$). The drop percentage ($D$) correlates with $\alpha$: a higher $\alpha$ means fewer inputs are dropped, leading to faster training but possibly lower test accuracy.

%% file: body/tables/models.tex
\resizebox{2\columnwidth}{!}{
\begin{tabular}{| l | l | c | c | c | c | c | c | c |}
\hline
\multirow{2}{*}{\textbf{Model}} & \multirow{2}{*}{\textbf{Dataset}} &  \multicolumn{2}{c|}{\textbf{Features}} &  \multicolumn{2}{c|}{\textbf{Neural Network Configuration}} &
\multicolumn{3}{c|}{\textbf{Embedding Tables Configuration}}
\\
\cline{3-9} & & \textbf{Dense} & \textbf{Sparse} & \textbf{Bottom MLP} & \textbf{Top MLP} & \textbf{No of Rows} & \textbf{Row Dim} & \textbf{Size}
\\
\hline
RM1 (TBSM) & Taobao Alibaba~\citep{alibaba} & 1 & 3 &  1-16 \& 22-15-15 & 30-60-1 & 5.1M & 16 & 0.3 GB \\
 \hline
RM2 (DLRM) & Criteo Kaggle~\citep{criteokaggle} & 13 & 26 & 13-512-256-64-16 & 512-256-1 & 33.8M & 16 & 2 GB  \\
 \hline
RM3 (DLRM) & Criteo Terabyte~\citep{criteoterabyte} & 13 & 26 & 13-512-256-64 & 512-512-256-1 & 266M  & 64 & 63 GB \\
 \hline
RM4 (DLRM) & Avazu~\citep{avazu} & 1 & 21 & 1-512-256-64-16 & 512-256-1 & 9.3M  & 16 & 0.55 GB \\
 \hline
\end{tabular}}

%% file: body/tables/server.tex
\begin{table}[h!]
\centering
\caption{Training System Specifications}
\resizebox{0.8\columnwidth}{!}{
\begin{tabular}{|c|c|c|c|}
 \hline
 \textbf{Device} & \textbf{Architecture} & \textbf{Memory} & \textbf{Storage}  \\
 \hline
 CPU & Intel Xeon & 192 GB & 1.9 TB \\
     & Silver 4116 (2.1GHz) & DDR4 (76.8GB/s) & NVMe SSD\\
 \hline
 GPU & Nvidia Tesla & 16 GB  &  - \\
 & V100 (1.2GHz) & HBM-2.0 (900GB/s) & \\
 \hline

\end{tabular}}
\label{tab:systemconfig}
\end{table}

%% file: body/tables/metric.tex
\begin{table}[h!]
\centering
\caption{Evaluation Metrics: Slipstream versus DLRM}
\resizebox{1\columnwidth}{!}{
\begin{tabular}{| l | c | c | c | c | c | c | }
\hline
\multirow{2}{*}{\textbf{Metric}} &  \multicolumn{2}{c|}{\textbf{RM2}} &  \multicolumn{2}{c|}{\textbf{RM3}} &
\multicolumn{2}{c|}{\textbf{RM4}}
\\
\cline{2-7} & \textbf{DLRM} & \textbf{Slipstream} & \textbf{DLRM} & \textbf{Slipstream} & \textbf{DLRM} & \textbf{Slipstream} \\
\hline
AUC  & 0.80 & \textbf{0.803} & 0.791 & \textbf{0.793} & 0.764 & \textbf{0.782}\\
 \hline
Accuracy (\%) & 78.63 & \textbf{78.86} & 81.16 & \textbf{81.23} & 83.59 & \textbf{83.77} \\
 \hline
BCE Loss & 	0.456 & \textbf{0.453} & 0.421 & \textbf{0.420} & 0.387 & \textbf{0.378}\\
 \hline
\end{tabular}}
\label{tab:evalmetrics}
\end{table}

%% file: body/tables/slipstream_ln.tex
\begin{table}[h!]
\centering
\caption{Ablation study comparing the effect of Slipstream accuracy recovery using LayerNorm}
\resizebox{1\columnwidth}{!}{
\begin{tabular}{| l | c | c | c | c | c | c | }
\hline
\multirow{3}{*}{\textbf{Metric}} &  \multicolumn{2}{c|}{\textbf{RM2}} &  \multicolumn{2}{c|}{\textbf{RM3}} &
\multicolumn{2}{c|}{\textbf{RM4}}
\\
\cline{2-7} & \textbf{Slipstream} & \textbf{Slipstream} & \textbf{Slipstream} & \textbf{Slipstream} & \textbf{Slipstream} & \textbf{Slipstream}
\\
 & \textbf{w/ LN} & \textbf{w/o LN} & \textbf{w/ LN} & \textbf{w/o LN} & \textbf{w/ LN} & \textbf{w/o LN} \\
\hline
AUC  & \textbf{0.803} & 0.797 & \textbf{0.793} & 0.787 & \textbf{0.782} & 0.757\\
 \hline
Accuracy (\%) & \textbf{78.86} & 78.59 & \textbf{81.23} & 81.13 & \textbf{83.77} & 83.56 \\
 \hline
BCE Loss & 	\textbf{0.453} & 0.458 & \textbf{0.420} & 0.422 & \textbf{0.378} & 0.389\\
 \hline
\end{tabular}}
\label{tab:layer_norm}
\end{table}

%% file: body/tables/alpha_drop_percentage.tex
\begin{table*}[h!]
\centering
\caption{Effect of Slipstream parameter $\alpha$ on drop percentage ($D$)}
\resizebox{0.6\textwidth}{!}{
\begin{tabular}{|l|c|c|c|c|c|c|c|c|c}
 \hline
 \textbf{$\alpha$} & \textbf{$\alpha$ = 0} & \textbf{$\alpha$ = 4} & \textbf{$\alpha$ = 8} & \textbf{$\alpha$ =12} & \textbf{$\alpha$ = 16} & \textbf{$\alpha$ = 20} & \textbf{$\alpha$ = 24} &
 \textbf{$\alpha$ = 26}  \\ 
 \hline
 RM1 (Taobao) & 100\% & 28.00\% & N/A &	N/A	 &N/A	& N/A &	N/A &	N/A	 \\
 \hline
 RM2 (Criteo Kaggle) & 100\%	& 28.99\% & 27.93\%	& 30.59\% & 30.57\%	& 28.08\% &	31.40\%	& 30.65\%	 \\
 \hline
 RM3 (Criteo Terabyte) & 100\%	& 28.27\% & 28.37\%	& 29.0\% & 26.67\%	& 26.86\% & 32.39\% & 0\% \\
 \hline
 RM4 (Avazu) & 100\%	& 27.48\% &	0\%	& 28.49\% & 27.1\% &	32.46\%	& N/A &	N/A	  \\
 \hline
\end{tabular}}
\label{tab:alpha2}
\end{table*}

%% file: body/related.tex
\section{Discussion}
\subsection{Slipstream versus Sampling-based Approaches}
Prior work such as FAE~\cite{fae} has employed input sampling to determine `hot' embedding. However, such works do not investigate the semantics of variable and invariable embedding entries. To our knowledge, Slipstream is the first work that uses sampling to identify and filter stale embeddings and their inputs. It is also crucial to note that techniques from FAE cannot be readily applied to identify stale embeddings. This limitation arises because prior work in this space has not recognized the existence of such stagnant `hot' embeddings.

\subsection{Slipstream in Commercial Environments}

Slipstream's advantages become even more evident in commercial settings where large recommender models are prevalent. In such expansive systems, the advantages of bypassing stale embeddings become more prominent, both within individual compute nodes and across the network. This is mainly attributable to Slipstream's considerable bandwidth and compute savings. Thus, Slipstream is orthogonal to other works and can be employed alongside others.

\subsection{Slipstream in GPU-only mode}

Certain commercial frameworks~\cite{hugectr, zionex-fb} employ a GPU-only approach for training extensive recommendation models, leveraging model parallelism across distributed GPU nodes. However, this configuration often encounters challenges due to bottlenecked all-to-all communication, stemming from bandwidth limitations in inter-node connections such as Infiniband or Ethernet. In contrast, Slipstream's strategy of bypassing stale inputs can alleviate the burden of all-to-all communication, thus expediting the training process.

\section{Prior Work}
We broadly address the relevant prior work in this category into two parts -- works on embedding placement and works on efficient training of recommender models.

\noindent \textbf{1. Optimized Embedding Placement:} Prior works have explored various embedding data placements for recommender models to mitigate the large memory footprint, curtailing performance. FAE~\citep{fae} determines semantically the access patterns of embedding entries. It places highly accessed embeddings close to computing on GPUs and the remainder of the embeddings on the CPU's main memory. Works in~\citep{acun2020understanding, baidu} determine the embedding placement strategy in heterogeneous system stacks.

\noindent \textbf{2. Efficient Recommender Model Training:} Past work~\citep{recnmp} is motivated by spatial and temporal reuse in embedding sparse table-related operations. They propose a custom near-memory processing design to increase the throughput. The notion of skew in terms of access has also been exploited by works ~\citep{MixDimTrick19, compemd} to redesign the embedding structure. Access skew has been further leveraged to develop compression mechanisms ~\citep{ttrec} and execution strategies ~\citep{bagpipe, recshard, scalable-reco} to accelerate training. The most recent work, Hotline~\citep{hotline}, uses an FPGA-based accelerator to perform parameter gathering for embeddings to pipeline inputs to the GPU while previous mini-batches are executing.

\noindent \textbf{3. Machine learning accelerators:} 
There are proposals for accelerators designed to execute the compute portion of deep learning models~\cite{tpu, tabla:hpca, cosmic:micro, maeri, minerva, eyeriss, brainwave}, including some for collaborative filtering-based recommender models~\cite{tabla:hpca, cosmic:micro, hotline, zionex-fb, zionex-paper, zion}. However, Slipstream can speed up training without specialized accelerators, though using them could offer additional benefits.

\noindent \textbf{4. Feature Interactions for Improved Convergence:} 
The feature interaction layer enhances recommendation tasks using complex information from sparse and dense features. Various methods model these cross-features differently. For instance, DLRM~\cite{dlrm} uses dot-product-based interactions, and DeepFM~\cite{DeepFM} introduces 2nd-order cross-features to an MLP for generating higher-order interactions implicitly. 

Other approaches include replacing DeepFM's MLP with more sophisticated networks, such as the deep cross network (DCN)~\cite{dcnv2} and xDeepFM~\cite{xDeepFM} for direct and intricate feature crosses, the inner and outer product network (PNN)~\cite{PNN} for product-based interactions, and self-attention mechanisms in AutoInt~\cite{autoint}. Ad-Rec~\cite{adrec} employs masked transformer-based feature interactions for faster convergence of recommendation models. Slipstream, however, does not rely on changing the model but employs skipping the inputs by identifying stale embeddings and would work on top of any prior state-of-the-art feature interaction techniques.

\noindent \textbf{5. Early stopping for DNNs:} Previous research has explored early stopping techniques for deep neural networks in fields such as image recognition, machine translation, and natural language processing~\cite{heckel2020early, bai2021understanding, ji2021early}.
Most such techniques have looked into early stopping in case of no validation set~\cite{mahsereci2017early} or with validation set~\cite{prechelt1998automatic}.
Unlike these methods, Slipstream intelligently applies early stopping to training recommendation models by identifying and skipping inputs with stale parameters. Notably, the state-of-the-art works in recommender models \emph{do not} dive into the semantic-driven variability properties of embedding tables to ensure that only relevant training is being performed.

\section{Conclusions}
Recommendation models have grown due to increasing users and items, leading to higher communication and computational overheads during training. Previous work has explored approaches like hybrid CPU-GPU setups and leveraging semantics, such as embedding popularity, to enhance training throughput. In this paper, we propose Slipstream, a runtime framework that utilizes the variations in embedding values to identify and skip fully trained embeddings on the fly. Slipstream helps reduce communication and computational overheads during training. Slipstream achieves a training time reduction of approximately 2$\times$, 2.4$\times$, 1.2$\times$, and 1.175$\times$ compared to commercial XDL, Intel-Optimized DLRM, FAE, and hotline baselines, while also \textbf{gaining 0.06 testing accuracy} across these models and datasets.

%% file: main.bbl
\begin{thebibliography}{10}

\bibitem{hugectr}
{NVIDIA Merlin: HugeCTR}.
\newblock \url{https://github.com/NVIDIA-Merlin/HugeCTR}.

\bibitem{tensorflow}
Martín Abadi, Ashish Agarwal, Paul Barham, Eugene Brevdo, Zhifeng Chen, Craig Citro, Greg Corrado, Andy Davis, Jeffrey Dean, Matthieu Devin, Sanjay Ghemawat, Ian Goodfellow, Andrew Harp, Geoffrey Irving, Michael Isard, Yangqing Jia, Rafal Jozefowicz, Lukasz Kaiser, Manjunath Kudlur, Josh Levenberg, Dan Mané, Rajat Monga, Sherry Moore, Derek Murray, Chris Olah, Mike Schuster, Jonathon Shlens, Benoit Steiner, Ilya Sutskever, Kunal Talwar, Paul Tucker, Vincent Vanhoucke, Vijay Vasudevan, Fernanda Viégas, Oriol Vinyals, Pete Warden, Martin Wattenberg, Martin Wicke, Yuan Yu, and Xiaoqiang Zheng.
\newblock {TensorFlow: Large-Scale Machine Learning on Heterogeneous Distributed Systems}, 2015.

\bibitem{acun2020understanding}
Bilge Acun, Matthew Murphy, Xiaodong Wang, Jade Nie, Carole-Jean Wu, and Kim Hazelwood.
\newblock {Understanding Training Efficiency of Deep Learning Recommendation Models at Scale}, 2020.

\bibitem{fae}
Muhammad Adnan, Yassaman Ebrahimzadeh~Maboud, Divya Mahajan, and Prashant Nair.
\newblock {Accelerating Recommendation System Trainingby Leveraging Popular Choices.}
\newblock In {\em VLDB}, 2022.

\bibitem{hotline}
Muhammad Adnan, Yassaman~Ebrahimzadeh Maboud, Divya Mahajan, and Prashant~J. Nair.
\newblock Heterogeneous acceleration pipeline for recommendation system training, 2022.

\bibitem{adrec}
Muhammad Adnan, Yassaman~Ebrahimzadeh Maboud, Divya Mahajan, and Prashant~J. Nair.
\newblock Ad-rec: Advanced feature interactions to address covariate-shifts in recommendation networks, 2023.

\bibitem{bagpipe}
Saurabh Agarwal, Ziyi Zhang, and Shivaram Venkataraman.
\newblock Bagpipe: Accelerating deep recommendation model training, 2022.

\bibitem{alibaba}
Alibaba.
\newblock {User Behavior Data from Taobao for Recommendation}.
\newblock \url{https://tianchi.aliyun.com/dataset/dataDetail?dataId=649userId=1}.

\bibitem{bai2021understanding}
Yingbin Bai, Erkun Yang, Bo~Han, Yanhua Yang, Jiatong Li, Yinian Mao, Gang Niu, and Tongliang Liu.
\newblock Understanding and improving early stopping for learning with noisy labels.
\newblock {\em Advances in Neural Information Processing Systems}, 34:24392--24403, 2021.

\bibitem{eyeriss}
Yu-Hsin Chen, Joel Emer, and Vivienne Sze.
\newblock {Eyeriss: A Spatial Architecture for Energy-Efficient Dataflow for Convolutional Neural Networks}.
\newblock In {\em ISCA}, 2016.

\bibitem{brainwave}
Eric Chung, Jeremy Fowers, Kalin Ovtcharov, , Adrian Caulfield, Todd Massengill, Ming Liu, Mahdi Ghandi, Daniel Lo, Steve Reinhardt, Shlomi Alkalay, Hari Angepat, Derek Chiou, Alessandro Forin, Doug Burger, Lisa Woods, Gabriel Weisz, Michael Haselman, and Dan Zhang.
\newblock {Serving DNNs in Real Time at Datacenter Scale with Project Brainwave}.
\newblock {\em IEEE Micro}, 38:8--20, March 2018.

\bibitem{criteokaggle}
CriteoLabs.
\newblock {Criteo Display Ad Challenge}.
\newblock \url{https://www.kaggle.com/c/criteo-display-ad-challenge}.

\bibitem{criteoterabyte}
CriteoLabs.
\newblock {Terabyte Click Logs}.
\newblock \url{https://labs.criteo.com/2013/12/download-terabyte-click-logs}.

\bibitem{sparsemat}
Jeremy Fowers, Kalin Ovtcharov, Karin Strauss, Eric Chung, and Greg Stitt.
\newblock {A High Memory Bandwidth FPGA Accelerator for Sparse Matrix-Vector Multiplication}.
\newblock In {\em International Symposium on Field-Programmable Custom Computing Machines}. IEEE, May.

\bibitem{MixDimTrick19}
Antonio Ginart, Maxim Naumov, Dheevatsa Mudigere, Jiyan Yang, and James Zou.
\newblock Mixed dimension embeddings with application to memory-efficient recommendation systems.
\newblock {\em CoRR}, abs/1909.11810, 2019.

\bibitem{netflixreco}
Carlos~A. Gomez-Uribe and Neil Hunt.
\newblock The netflix recommender system: Algorithms, business value, and innovation.
\newblock {\em ACM Trans. Manage. Inf. Syst.}, 6(4), December 2016.

\bibitem{DeepFM}
Huifeng Guo, Ruiming Tang, Yunming Ye, Zhenguo Li, and Xiuqiang He.
\newblock Deepfm: a factorization-machine based neural network for ctr prediction.
\newblock {\em arXiv preprint arXiv:1703.04247}, 2017.

\bibitem{heckel2020early}
Reinhard Heckel and Fatih~Furkan Yilmaz.
\newblock Early stopping in deep networks: Double descent and how to eliminate it.
\newblock {\em arXiv preprint arXiv:2007.10099}, 2020.

\bibitem{tbsm}
T.~Ishkhanov, M.~Naumov, X.~Chen, Y.~Zhu, Y.~Zhong, A.~G. Azzolini, C.~Sun, F.~Jiang, A.~Malevich, and L.~Xiong.
\newblock {Time-based Sequence Model for Personalization and Recommendation Systems}.
\newblock {\em CoRR}, abs/2008.11922, 2020.

\bibitem{ji2021early}
Ziwei Ji, Justin Li, and Matus Telgarsky.
\newblock Early-stopped neural networks are consistent.
\newblock {\em Advances in Neural Information Processing Systems}, 34:1805--1817, 2021.

\bibitem{xdl}
Biye Jiang, Chao Deng, Huimin Yi, Zelin Hu, Guorui Zhou, Yang Zheng, Sui Huang, Xinyang Guo, Dongyue Wang, Yue Song, Liqin Zhao, Zhi Wang, Peng Sun, Yu~Zhang, Di~Zhang, Jinhui Li, Jian Xu, Xiaoqiang Zhu, and Kun Gai.
\newblock {XDL: An Industrial Deep Learning Framework for High-Dimensional Sparse Data}.
\newblock DLP-KDD '19, New York, NY, USA, 2019. Association for Computing Machinery.

\bibitem{tpu}
Norman~P. Jouppi, Cliff Young, Nishant Patil, David Patterson, Gaurav Agrawal, Raminder Bajwa, Sarah Bates, Suresh Bhatia, Nan Boden, Al~Borchers, Rick Boyle, Pierre-luc Cantin, Clifford Chao, Chris Clark, Jeremy Coriell, Mike Daley, Matt Dau, Jeffrey Dean, Ben Gelb, Tara~Vazir Ghaemmaghami, Rajendra Gottipati, William Gulland, Robert Hagmann, C.~Richard Ho, Doug Hogberg, John Hu, Robert Hundt, Dan Hurt, Julian Ibarz, Aaron Jaffey, Alek Jaworski, Alexander Kaplan, Harshit Khaitan, Daniel Killebrew, Andy Koch, Naveen Kumar, Steve Lacy, James Laudon, James Law, Diemthu Le, Chris Leary, Zhuyuan Liu, Kyle Lucke, Alan Lundin, Gordon MacKean, Adriana Maggiore, Maire Mahony, Kieran Miller, Rahul Nagarajan, Ravi Narayanaswami, Ray Ni, Kathy Nix, Thomas Norrie, Mark Omernick, Narayana Penukonda, Andy Phelps, Jonathan Ross, Matt Ross, Amir Salek, Emad Samadiani, Chris Severn, Gregory Sizikov, Matthew Snelham, Jed Souter, Dan Steinberg, Andy Swing, Mercedes Tan, Gregory Thorson, Bo~Tian, Horia Toma, Erick Tuttle, Vijay
  Vasudevan, Richard Walter, Walter Wang, Eric Wilcox, and Doe~Hyun Yoon.
\newblock {In-Datacenter Performance Analysis of a Tensor Processing Unit}.
\newblock In {\em Proceedings of the 44th Annual International Symposium on Computer Architecture}, ISCA '17, page 1–12, New York, NY, USA, 2017. Association for Computing Machinery.

\bibitem{avazu}
Kaggle.
\newblock {Avazu mobile ads CTR}.
\newblock \url{https://www.kaggle.com/c/avazu-ctr-prediction}.

\bibitem{dlrm-intel}
Dhiraj Kalamkar, Evangelos Georganas, Sudarshan Srinivasan, Jianping Chen, Mikhail Shiryaev, and Alexander Heinecke.
\newblock {Optimizing Deep Learning Recommender Systems Training on CPU Cluster Architectures}.
\newblock In {\em Proceedings of the International Conference for High Performance Computing, Networking, Storage and Analysis}, SC '20. IEEE Press, 2020.

\bibitem{recnmp}
L.~{Ke}, U.~{Gupta}, B.~Y. {Cho}, D.~{Brooks}, V.~{Chandra}, U.~{Diril}, A.~{Firoozshahian}, K.~{Hazelwood}, B.~{Jia}, H.~S. {Lee}, M.~{Li}, B.~{Maher}, D.~{Mudigere}, M.~{Naumov}, M.~{Schatz}, M.~{Smelyanskiy}, X.~{Wang}, B.~{Reagen}, C.~{Wu}, M.~{Hempstead}, and X.~{Zhang}.
\newblock {RecNMP: Accelerating Personalized Recommendation with Near-Memory Processing}.
\newblock In {\em 2020 ACM/IEEE 47th Annual International Symposium on Computer Architecture (ISCA)}, pages 790--803, 2020.

\bibitem{maeri}
Hyoukjun Kwon, Ananda Samajdar, and Tushar Krishna.
\newblock {MAERI: Enabling Flexible Dataflow Mapping over DNN Accelerators via Reconfigurable Interconnects}.
\newblock In {\em Proceedings of the Twenty-Third International Conference on Architectural Support for Programming Languages and Operating Systems}, ASPLOS '18, page 461–475, New York, NY, USA, 2018. Association for Computing Machinery.

\bibitem{xDeepFM}
Jianxun Lian, Xiaohuan Zhou, Fuzheng Zhang, Zhongxia Chen, Xing Xie, and Guangzhong Sun.
\newblock Xdeepfm: Combining explicit and implicit feature interactions for recommender systems.
\newblock In {\em Proceedings of the 24th ACM SIGKDD International Conference on Knowledge Discovery and Data Mining}, KDD '18, page 1754–1763, New York, NY, USA, 2018. Association for Computing Machinery.

\bibitem{tabla:hpca}
Divya Mahajan, Jongse Park, Emmanuel Amaro, Hardik Sharma, Amir Yazdanbakhsh, Joon Kim, and { Hadi Esmaeilzadeh}.
\newblock {\sc Tabla}: {A} unified template-based framework for accelerating statistical machine learning.
\newblock March 2016.

\bibitem{mahsereci2017early}
Maren Mahsereci, Lukas Balles, Christoph Lassner, and Philipp Hennig.
\newblock Early stopping without a validation set.
\newblock {\em arXiv preprint arXiv:1703.09580}, 2017.

\bibitem{zionex-fb}
meta.
\newblock {Meta recommender model training on ZionEX devices}.
\newblock \url{https://www.infoq.com/news/2021/05/facebook-zionex-training/}.

\bibitem{scalable-reco}
Dheevatsa Mudigere, Yuchen Hao, Jianyu Huang, Zhihao Jia, Andrew Tulloch, Srinivas Sridharan, Xing Liu, Mustafa Ozdal, Jade Nie, Jongsoo Park, Liang Luo, Jie~(Amy) Yang, Leon Gao, Dmytro Ivchenko, Aarti Basant, Yuxi Hu, Jiyan Yang, Ehsan~K. Ardestani, Xiaodong Wang, Rakesh Komuravelli, Ching-Hsiang Chu, Serhat Yilmaz, Huayu Li, Jiyuan Qian, Zhuobo Feng, Yinbin Ma, Junjie Yang, Ellie Wen, Hong Li, Lin Yang, Chonglin Sun, Whitney Zhao, Dimitry Melts, Krishna Dhulipala, KR~Kishore, Tyler Graf, Assaf Eisenman, Kiran~Kumar Matam, Adi Gangidi, Guoqiang~Jerry Chen, Manoj Krishnan, Avinash Nayak, Krishnakumar Nair, Bharath Muthiah, Mahmoud khorashadi, Pallab Bhattacharya, Petr Lapukhov, Maxim Naumov, Ajit Mathews, Lin Qiao, Mikhail Smelyanskiy, Bill Jia, and Vijay Rao.
\newblock Software-hardware co-design for fast and scalable training of deep learning recommendation models.
\newblock In {\em Proceedings of the 49th Annual International Symposium on Computer Architecture}, ISCA '22, page 993–1011, New York, NY, USA, 2022. Association for Computing Machinery.

\bibitem{zionex-paper}
Dheevatsa Mudigere, Yuchen Hao, Jianyu Huang, Andrew Tulloch, Srinivas Sridharan, Xing Liu, Mustafa Ozdal, Jade Nie, Jongsoo Park, Liang Luo, Jie~Amy Yang, Leon Gao, Dmytro Ivchenko, Aarti Basant, Yuxi Hu, Jiyan Yang, Ehsan~K. Ardestani, Xiaodong Wang, Rakesh Komuravelli, Ching{-}Hsiang Chu, Serhat Yilmaz, Huayu Li, Jiyuan Qian, Zhuobo Feng, Yinbin Ma, Junjie Yang, Ellie Wen, Hong Li, Lin Yang, Chonglin Sun, Whitney Zhao, Dimitry Melts, Krishna Dhulipala, K.~R. Kishore, Tyler Graf, Assaf Eisenman, Kiran~Kumar Matam, Adi Gangidi, Guoqiang~Jerry Chen, Manoj Krishnan, Avinash Nayak, Krishnakumar Nair, Bharath Muthiah, Mahmoud khorashadi, Pallab Bhattacharya, Petr Lapukhov, Maxim Naumov, Lin Qiao, Mikhail Smelyanskiy, Bill Jia, and Vijay Rao.
\newblock High-performance, distributed training of large-scale deep learning recommendation models.
\newblock {\em CoRR}, abs/2104.05158, 2021.

\bibitem{zion}
Dheevatsa Mudigere and Whitney Zhao.
\newblock Hw/sw co-design for future ai platforms - large memory unified training platform (zion), 2019.

\bibitem{dlrm}
Maxim Naumov, Dheevatsa Mudigere, Hao{-}Jun~Michael Shi, Jianyu Huang, Narayanan Sundaraman, Jongsoo Park, Xiaodong Wang, Udit Gupta, Carole{-}Jean Wu, Alisson~G. Azzolini, Dmytro Dzhulgakov, Andrey Mallevich, Ilia Cherniavskii, Yinghai Lu, Raghuraman Krishnamoorthi, Ansha Yu, Volodymyr Kondratenko, Stephanie Pereira, Xianjie Chen, Wenlin Chen, Vijay Rao, Bill Jia, Liang Xiong, and Misha Smelyanskiy.
\newblock {Deep Learning Recommendation Model for Personalization and Recommendation Systems}.
\newblock {\em CoRR}, abs/1906.00091, 2019.

\bibitem{nccl}
Nvidia.
\newblock {{NVIDIA Collective Communications Library (NCCL)}}.
\newblock \url{https://docs.nvidia.com/deeplearning/nccl/index.html}.

\bibitem{nvlink}
Nvidia.
\newblock Nvlink.
\newblock \url{https://www.nvidia.com/en-us/data-center/nvlink/}.

\bibitem{cosmic:micro}
Jongse Park, Hardik Sharma, Divya Mahajan, Joon~Kyung Kim, Preston Olds, and { Hadi Esmaeilzadeh}.
\newblock {Scale-Out Acceleration for Machine Learnng}.
\newblock October 2017.

\bibitem{pytorch}
Adam Paszke, Sam Gross, Soumith Chintala, Gregory Chanan, Edward Yang, Zachary DeVito, Zeming Lin, Alban Desmaison, Luca Antiga, and Adam Lerer.
\newblock {Automatic differentiation in PyTorch}.
\newblock 2017.

\bibitem{prechelt1998automatic}
Lutz Prechelt.
\newblock Automatic early stopping using cross validation: quantifying the criteria.
\newblock {\em Neural networks}, 11(4):761--767, 1998.

\bibitem{PNN}
Yanru Qu, Han Cai, Kan Ren, Weinan Zhang, Yong Yu, Ying Wen, and Jun Wang.
\newblock Product-based neural networks for user response prediction.
\newblock In {\em 2016 IEEE 16th international conference on data mining (ICDM)}, pages 1149--1154. IEEE, 2016.

\bibitem{minerva}
B.~{Reagen}, P.~{Whatmough}, R.~{Adolf}, S.~{Rama}, H.~{Lee}, S.~K. {Lee}, J.~M. {Hernández-Lobato}, G.~{Wei}, and D.~{Brooks}.
\newblock {Minerva: Enabling Low-Power, Highly-Accurate Deep Neural Network Accelerators}.
\newblock In {\em 2016 ACM/IEEE 43rd Annual International Symposium on Computer Architecture (ISCA)}, pages 267--278, June 2016.

\bibitem{recshard}
Geet Sethi, Bilge Acun, Niket Agarwal, Christos Kozyrakis, Caroline Trippel, and Carole-Jean Wu.
\newblock Recshard: Statistical feature-based memory optimization for industry-scale neural recommendation, 2022.

\bibitem{compemd}
Hao-Jun~Michael Shi, Dheevatsa Mudigere, Maxim Naumov, and Jiyan Yang.
\newblock {\em {Compositional Embeddings Using Complementary Partitions for Memory-Efficient Recommendation Systems}}, page 165–175.
\newblock Association for Computing Machinery, New York, NY, USA, 2020.

\bibitem{amazonreco}
B.~{Smith} and G.~{Linden}.
\newblock Two decades of recommender systems at amazon.com.
\newblock {\em IEEE Internet Computing}, 21(3):12--18, 2017.

\bibitem{autoint}
Weiping Song, Chence Shi, Zhiping Xiao, Zhijian Duan, Yewen Xu, Ming Zhang, and Jian Tang.
\newblock Autoint: Automatic feature interaction learning via self-attentive neural networks.
\newblock In {\em Proceedings of the 28th ACM International Conference on Information and Knowledge Management}, pages 1161--1170, 2019.

\bibitem{compressreco}
Yang Sun, Fajie Yuan, Min Yang, Guoao Wei, Zhou Zhao, and Duo Liu.
\newblock {A Generic Network Compression Framework for Sequential Recommender Systems}.
\newblock In {\em Proceedings of the 43rd International ACM SIGIR Conference on Research and Development in Information Retrieval}, SIGIR '20, page 1299–1308, New York, NY, USA, 2020. Association for Computing Machinery.

\bibitem{dcnv2}
Ruoxi Wang, Rakesh Shivanna, Derek Cheng, Sagar Jain, Dong Lin, Lichan Hong, and Ed~Chi.
\newblock Dcn v2: Improved deep \& cross network and practical lessons for web-scale learning to rank systems.
\newblock In {\em Proceedings of the web conference 2021}, pages 1785--1797, 2021.

\bibitem{facebook:ml}
C.~{Wu}, D.~{Brooks}, K.~{Chen}, D.~{Chen}, S.~{Choudhury}, M.~{Dukhan}, K.~{Hazelwood}, E.~{Isaac}, Y.~{Jia}, B.~{Jia}, T.~{Leyvand}, H.~{Lu}, Y.~{Lu}, L.~{Qiao}, B.~{Reagen}, J.~{Spisak}, F.~{Sun}, A.~{Tulloch}, P.~{Vajda}, X.~{Wang}, Y.~{Wang}, B.~{Wasti}, Y.~{Wu}, R.~{Xian}, S.~{Yoo}, and P.~{Zhang}.
\newblock Machine learning at facebook: Understanding inference at the edge.
\newblock In {\em 2019 IEEE International Symposium on High Performance Computer Architecture (HPCA)}, pages 331--344, Feb 2019.

\bibitem{compressreco2}
Xiaorui Wu, Hong Xu, Honglin Zhang, Huaming Chen, and Jian Wang.
\newblock {Saec: similarity-aware embedding compression in recommendation systems}.
\newblock In {\em Proceedings of the 11th ACM SIGOPS Asia-Pacific Workshop on Systems}, pages 82--89, 2020.

\bibitem{ttrec}
Chunxing Yin, Bilge Acun, Xing Liu, and Carole-Jean Wu.
\newblock {TT-Rec: Tensor Train Compression for Deep Learning Recommendation Models}, 2021.

\bibitem{baidu}
Weijie Zhao, Deping Xie, Ronglai Jia, Yulei Qian, Ruiquan Ding, Mingming Sun, and Ping Li.
\newblock Distributed hierarchical gpu parameter server for massive scale deep learning ads systems, 2020.

\bibitem{din}
Guorui Zhou, Xiaoqiang Zhu, Chenru Song, Ying Fan, Han Zhu, Xiao Ma, Yanghui Yan, Junqi Jin, Han Li, and Kun Gai.
\newblock Deep interest network for click-through rate prediction.
\newblock In {\em Proceedings of the 24th ACM SIGKDD international conference on knowledge discovery \& data mining}, pages 1059--1068, 2018.

\end{thebibliography}
